\documentclass[journal]{ieeetran}
\usepackage{tikz}
\usepackage{amsmath}
\usepackage{hyperref}
\usepackage{array}
\usepackage{wrapfig}
\usepackage{multirow}
\usepackage{xcolor}
\usepackage{subcaption}
\usepackage{authblk}
\usepackage{pdfpages}
\usepackage{xparse}
\usepackage{tabu}
\usepackage{wrapfig}
\usepackage{xcolor}
\usepackage{soul}
\usepackage{pgfplots}
\usepackage{amssymb}
\usepackage{flexisym}

\usepackage{amsmath}
\usepackage{mathtools}
\usepackage{xcolor}
\usepackage{lipsum} 
\usepackage[colorinlistoftodos]{todonotes}
\usepackage{pdfcomment}

\usepackage[T1]{fontenc}
\DeclareFontFamily{T1}{calligra}{}
\DeclareFontShape{T1}{calligra}{m}{n}{<->s*[1.44]callig15}{}
\DeclareMathAlphabet\mathcalligra   {T1}{calligra} {m} {n}
\DeclareMathAlphabet\mathzapf       {T1}{pzc} {mb} {it}

\newcommand{\norm}[1]{\left\lVert#1\right\rVert}

\makeatletter
\newcommand*{\rom}[1]{\expandafter\@slowromancap\romannumeral #1@}
\makeatother

\title{Esophageal Tumor Segmentation in CT Images using Dilated Dense Attention Unet (DDAUnet) }

\author[1]{Sahar~Yousefi}
\author[1]{Hessam~Sokooti}
\author[1]{Mohamed~S.~Elmahdy}
\author[2]{Irene~M.~Lips}
\author[3]{Mohammad~T.~Manzuri~Shalmani}
\author[2,4]{Roel~T.~Zinkstok}
\author[2]{Frank~J.W.M.~Dankers}
\author[1,2,5]{Marius~Staring}

\affil[1]{Department of Radiology, Leiden University Medical Center, Leiden, The Netherlands}
\affil[2]{Department of Radiotherapy, Leiden University Medical Center, Leiden, The Netherlands}
\affil[3]{Department of Computer Engineering, Sharif University of Technology, Tehran, Iran}
\affil[4]{Department of Radiation Oncology, Netherlands Cancer Institute, Amsterdam, The Netherlands}
\affil[5]{Delft University of Technology, Delft, The Netherlands}

\affil[ ]{\textit {s.yousefi.radi@lumc.nl}}

\begin{document}
\maketitle
%

%

\begin{abstract}
Manual or automatic delineation of the esophageal tumor in CT images is known to be very challenging. This is due to the low contrast between the tumor and adjacent tissues, the anatomical variation of the esophagus, as well as the occasional presence of foreign bodies (e.g. feeding tubes). Physicians therefore usually exploit additional knowledge such as endoscopic findings, clinical history, additional imaging modalities like PET scans. Achieving his additional information is time-consuming, while the results are error-prone and might lead to non-deterministic results. In this paper we aim to investigate if and to what extent a simplified clinical workflow based on CT alone, allows one to automatically segment the esophageal tumor with sufficient quality. For this purpose, we present a fully automatic end-to-end esophageal tumor segmentation method based on convolutional neural networks (CNNs). The proposed network, called Dilated Dense Attention Unet (DDAUnet), leverages spatial and channel attention gates in each dense block to selectively concentrate on determinant feature maps and regions. Dilated convolutional layers are used to manage GPU memory and increase the network receptive field. We collected a dataset of 792 scans from 288 distinct patients including varying anatomies with \mbox{air pockets}, feeding tubes and proximal tumors. 
Repeatability and reproducibility studies were conducted for three distinct splits of training and validation sets. The proposed network achieved a $\mathrm{DSC}$ value of $0.79 \pm 0.20$, a mean surface distance of $5.4 \pm 20.2mm$ and $95\%$ Hausdorff distance of $14.7 \pm 25.0mm$ for 287 test scans, demonstrating promising results with a simplified clinical workflow based on CT alone. Our code is publicly available via \url{https://github.com/yousefis/DenseUnet_Esophagus_Segmentation}.

\end{abstract}


\section{Introduction}

\begin{figure*}[!tb]
\centering
\setlength\extrarowheight{-3pt}
\input{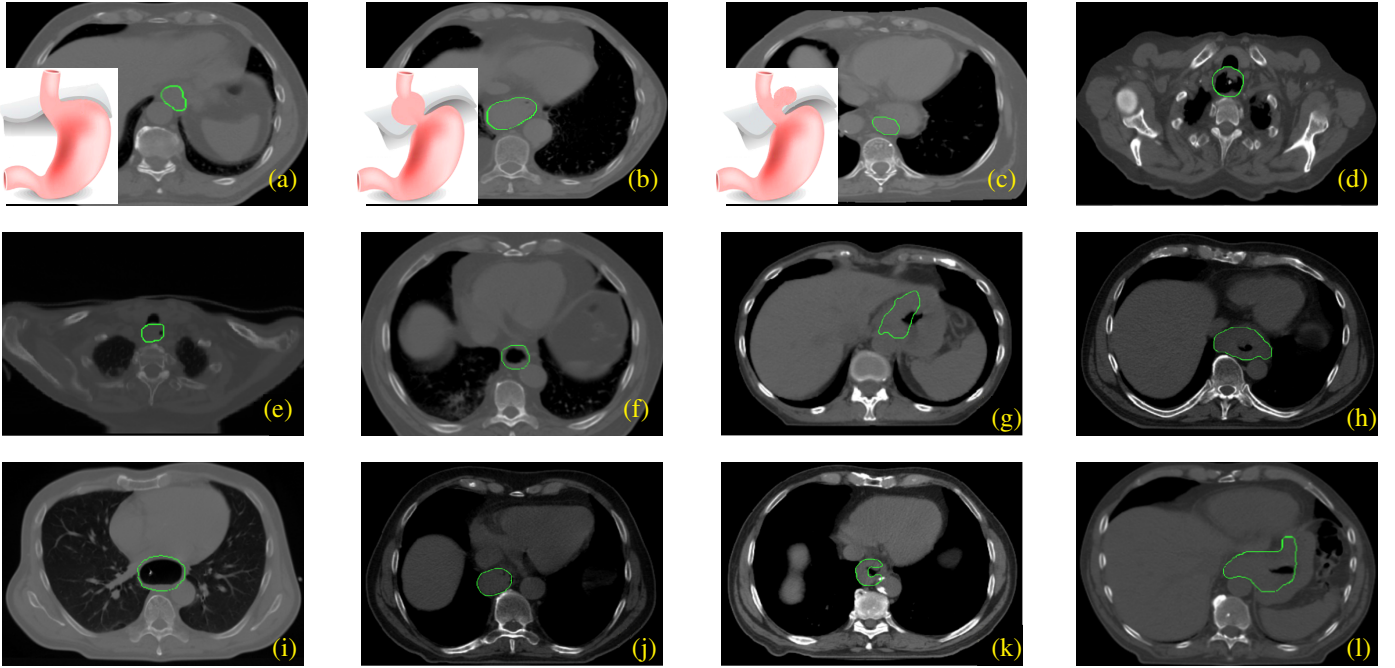}
\caption{Variations in shape and location of the tumor. Green contours show the manual delineation of the GTVs. (a) normal junction of esophagus and stomach, (b) hiatal hernia (type I): migration of esophagogastric junction through the gap in the cranial direction, (c) hiatal hernia (type II): migration of esophagogastric junction in the chest, (d) proximal tumor including an \mbox{air pocket} and feeding tube, (e) proximal tumor including an \mbox{air pocket}, (f) tumor including an \mbox{air pocket}, (g,h) junction tumor (extension of the tumor into the stomach) including an \mbox{air pocket}, (i) tumor including an \mbox{air pocket} and feeding tube, (j) relocation of the esophagus to the left of aorta, (k) a variety in the shape of the tumor, (l) junction tumor including an \mbox{air pocket}.}\label{fig:variety_cases}
\end{figure*}

Esophageal cancer is one of the least studied cancers \cite{enzinger2003esophageal}, while it is lethal in most patients \cite{ferlay2015cancer}. Because of the very poor survival rate three standard treatment options are available, i.e. chemoradiotherapy (CRT), neoadjuvant CRT followed by surgical resection, or radical radiotherapy \cite{mamede2007fdg}. For this purpose, rapid and accurate delineation of the target volume in CT images plays a very important role in therapy and disease control. 
The complexities raised by automatic esophageal tumor delineation in CT images can be divided into several categories: textural similarities and the absence of contrast between the tumor and its adjacent tissues; the anatomical variation of different patients either intrinsically or caused by a disease, like a hiatal hernia in which part of the stomach bulges into the chest cavity through an opening of the diaphragm (see Figure \ref{fig:variety_cases}-(e) and (i)), extension of tumor into the stomach, or existence of \mbox{air pocket} inside the esophagus; existence of foreign bodies during the treatment, like a feeding tube or surgical clips inside the esophageal lumen. Figure \ref{fig:variety_cases} illustrates some of the challenging cases. These difficulties lead to a high degree of uncertainty associated with the target volume of the tumor, especially at the cranial and caudal border of the tumor \cite{nowee2019gross}. In order to overcome these complexities, physicians integrate CT imaging with the clinical history, endoscopic findings, endoscopic ultrasound, and other imaging modalities such as positron-emission tomography (PET) \cite{charles2009esophageal}. 
Obtaining these additional modalities is however a time-consuming and expensive process. Moreover, the process of manual delineation is a repetitive and tedious task, and often there is a lack of consensus on how to best segment the tumor from normal tissue. Despite using additional modalities and expert knowledge, the process of manual delineation still remains an ill-posed problem \cite{rousson2006probabilistic}. Nowee \textit{et al.} \cite{nowee2019gross} assessed manual delineation variability of gross tumor volume (GTV) between using CT and combined F-fluorodeoxyglucose PET (FDG-PET) \cite{lever2014quantification} and CT in esophageal cancer patients in a multi-institutional study by 20 observers. They concluded that the use of PET images can significantly influence the delineated volume in some cases, however its impact on observer variation is limited.

In this paper we aim to investigate if a simplified clinical workflow based on CT scans alone allows to automatically delineate the esophageal GTV with acceptable quality. Recently, there has been a revived interest in automating this process for both the esophagus and the esophageal tumor based on CT images alone \cite{jin2019deep, xue2017fully, liang2020auto}. 
Our earlier work \cite{yousefi2018esophageal}, leveraged the idea of dense blocks proposed by \cite{huang2017densely}, arranging them in a typical U-shape. In that study, the proposed method was trained and tested on 553 chests CT scans from 49 distinct patients and achieved a $\mathrm{DSC}$ value of $0.73 \pm 0.20$, and a 95$\%$ mean surface distance ($\mathrm{MSD}$) of $3.07 \pm 1.86$ mm for the test scans. Eight of the 85 scans in the test set had a $\mathrm{DSC}$ value lower than $0.50$, caused by the presence of air cavities and foreign bodies in the GTV, which was rarely seen in the training data. In order to enhance the robustness of the network, in the present study we extended that work. The main contributions of this paper are as follows:

\begin{enumerate}
\item We propose an end-to-end CNN for esophageal GTV segmentation on CT scans. Different from much of the previous work, which addressed segmentation of the esophagus itself, we focus on the more challenging tumor area (GTV). The proposed method is end-to-end, without intricate pre- or post-processing steps, and uses no information in addition to the CT scans;
\item We introduce dilated dense attention blocks which leverage spatial and channel attention to emphasize on the GTV related features. Also, dilation layers are used to support an exponential expansion of the receptive field and decrease the size of the network without loss of resolution;
\item We collected a dataset of $228$ distinct patients ($792$ scans). The dataset includes different varieties of anatomies, and presence of foreign bodies and \mbox{air pockets} in the esophageal lumen. In this study, all patients received either Neoadjuvant or definitive chemoradiotherapy treatment options. To the best of our knowledge, none of the related works have addressed such a comprehensive dataset.
\end{enumerate}

The initial results of this work were presented in \cite{yousefi2018esophageal}. The current paper includes a larger and more diverse dataset, and more elaborate evaluation. Also, we leverage dilated convolutional layers and attention gates in order to increase the receptive field and filter tumor relevant features.

\section{Related work}
Most automatic esophagus segmentation approaches have used either a shape or appearance model to guide the segmentation, where training such a guidance model is complicated. Rousson \textit{et al.} proposed a two-stage probabilistic shortest path approach to segment the esophagus from 2D CT images \cite{rousson2006probabilistic}. In the first stage, the aorta and left atrium are segmented and then registered to reference shapes in order to find a region of interest (ROI). In the second stage, the optimal esophagus centerline is extracted using the shortest path algorithm. Fieselmann \textit{et al.} proposed an automatic approach for segmenting the esophagus by detecting the air cavities that often constitute the esophagus \cite{fieselmann2008automatic}. For reducing the time complexity they confined all the computations to an ROI. Also, they proposed another method based on spatially-constrained shape interpolation in order to segment the esophagus in 2D CT images \cite{fieselmann2008esophagus}. In that investigation, two assumptions are considered: the shape of the esophagus changes smoothly, and there is no intersection between the esophagus and the other organs. 

In \cite{feulner2011probabilistic} a multi-step approach based on probabilistic models has been proposed to segment the esophagus on 3D CT scans. In that work, a pre-processing step is used to extract an ROI. Then, a discriminative learning technique is applied to label the voxels. In \cite{huang2006semi} an optical flow approach for semi-automated segmentation of CT images is used, where manually drawn curvature points are extended to contours by Fourier interpolation and afterwards, optical flow is used for registering the original contour to the other slices. This method is not only highly user-interactive but it also fails when the region to contour is topologically different between two slices. Feulner \textit{et al.} proposed a multi-step approach based on probabilistic models for automatic segmentation of the esophagus in 3D CT scans \cite{feulner2009fast}. In that work, by running a discriminative model for each axial slice, a set of approximated esophagus contours is extracted. Then, the contours are clustered and merged and afterwards, a Markov chain model is used for finding the most probable path through the axial slices. Ultimately, another discriminative model is used for refining the result. This approach just works for a manually selected ROI. The manual selection of an ROI was later extended to automatic ROI detection by a salient landmark on the chest \cite{feulner2010model}.

Kurugol \textit{et al.} presented a 3D level set model for segmenting the esophagus over the entire thoracic range employing a shape model, with a global and a locally deformable component \cite{kurugol2010locally}. In their work, an initial centerline estimation is required where an ad-hoc centerline estimator was used, which was only performed at the ROI of some predefined anatomical landmarks followed by interpolation for the remaining slices. Later, they extended their work by using prior spatial and appearance models estimated from the training set instead of using the ad-hoc estimator \cite{kurugol2011centerline}. 
In \cite{yang2017atlas}, a two-phase online atlas-based approach was proposed to rank and select a subset of optimal atlas candidates for segmentation of the esophagus on CT scans. Atlas-based approaches face some restrictions including the selection of optimal atlases and a correct representation of the study population.

Deep learning for medical image analysis has aroused broad attention in recent years \cite{yousefi2021asl, pezzotti2020adaptive, yousefi2019fast, elmahdy2018evaluation, elmahdy2019robust}. However, this technique has been limitedly used for esophagus segmentation and even less for esophageal tumor segmentation. In \cite{fechter20173d}, a fully convolutional neural network (FCNN) for segmenting the esophagus on 3D CT was proposed, surrounded the bottom-most of the heart and the topmost of the stomach. For refining the results, an active contour model and a random walker were used as post-processing steps. In that study, 50 scans were used as the training set and 20 scans as the test set. An average $\mathrm{DSC}$ value of \mbox{0.76 $\pm$ 0.11} for the test set was reported. A semi-automatic two-stage FCNN for 2D esophagus segmentation has been proposed by Trullo \textit{et al.} \cite{trullo2017fully}. The first stage performs a multi-organ segmentation in order to extract an ROI including the esophagus. Then the manually cropped ROI is fed to the second network to segment the esophagus. A $\mathrm{DSC}$ value of \mbox{0.72 $\pm$ 0.07} has been reported for this network with 25 scans as the training set and 30 scans as the testing set. For extracting the largest possible tumor region in 2D CT scans an FCNN was used by Hao \textit{et al.} \cite{hao2017esophagus}. Then they applied a graph cut for segmenting the tumor. They reported an average $\mathrm{DSC}$ value of \mbox{0.75 $\pm$ 0.04} for the four patients in the test set. Jin \textit{et al.} \cite{jin2019deep} introduced a spatial-context encoded deep esophageal clinical target volume (CTV) delineation framework to produce superior margin-based CTV boundaries. That work in an expensive pre-processing step encodes spatial context by computing the signed distance transform maps (SDMs) of the GTV, lymph nodes (LNs) and organs at risks (OARs) and then feeds the results with the CT image into a 3D CNN. In another work Jin \textit{et al.} \cite{jin2019accurate} proposed a two-stream chained 3D CNN fusion pipeline to segment esophageal GTVs using both CT and PET+CT scans. They evaluated their approach by conducting a 5-fold cross validation on scans of 110 patients. They reported that using PET images as complementary information can improve the $\mathrm{DSC}$ score from $0.73 \pm 0.16$ to $0.76 \pm 0.13$. Although reasonable results can be obtained in the approaches mentioned earlier, the problem of esophageal GTV segmentation in CT modalities without extra knowledge constraining the problem is known as an ill-posed problem and remains challenging \cite{xue2017fully}. 

Most of the mentioned works addressed esophagus segmentation and not esophageal tumor segmentation. However, esophageal tumor segmentation is a more complicated task due to the poor contrast of the tumor with respect to its adjacent tissues. It is especially difficult to define the start and end of the tumor without additional information such as endoscopic findings.

\section{The proposed method}
\subsection{Network architecture}
\begin{figure*}[ht]
  \centering
    \includegraphics[width=16.5cm,clip]{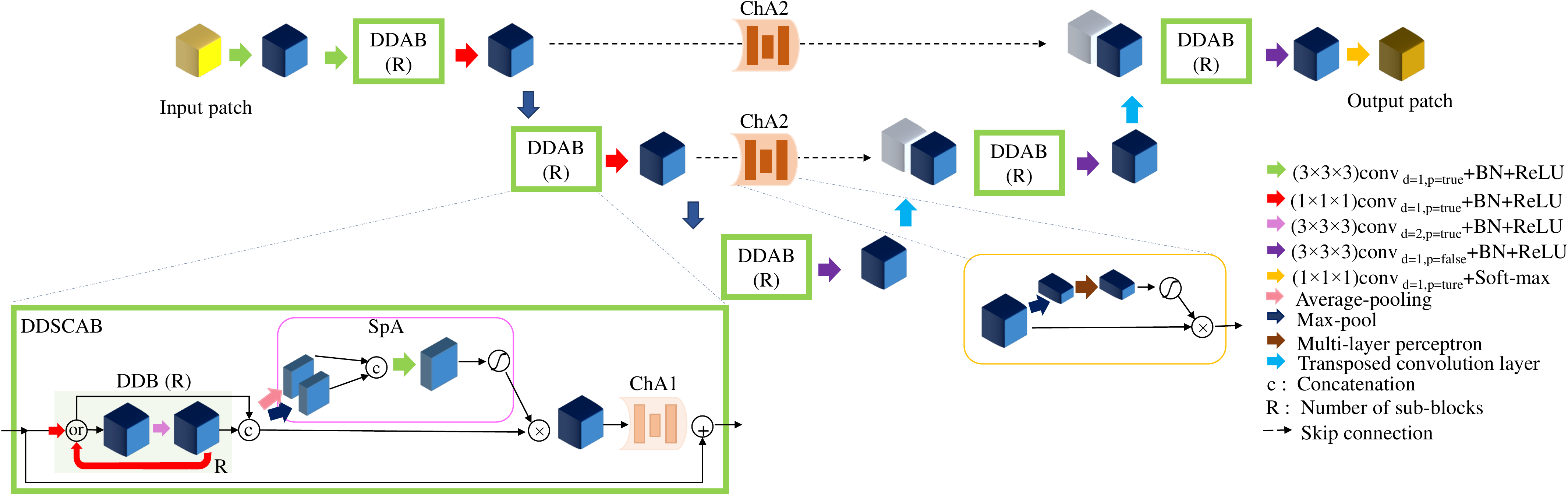}
  \caption{The architecture of the proposed method. DDSCAB and DDB stand for dilated dense spatial and channel attention block and dilated dense block, respectively. $R$ is the number of sub-DDBs. ChA1, ChA2, and SpA denote channel attention gate located on skip connections, channel attention inside the DDSCAB block, and spatial attention gates inside DDSCAB. ChA1, shown transparently here, is not included in the final network (DDAUnet), but is used in some of the experiments.}   
  \label{fig:architecture} 
\end{figure*}

Figure \ref{fig:architecture} shows a schematic of the proposed network, dubbed dilated dense attention Unet (DDAUnet). The network is composed of three levels, a down-sampling path for extracting contextual features and an up-sampling path for retrieving the lost resolution during extraction. In each level, different from our prior work \cite{yousefi2018esophageal}, we used dilated dense spatial and channel attention blocks (DDSCAB) which is composed of a dilated dense block (DDB) and a spatial attention gate and a channel attention gate which are denoted SpA and ChA1. Using loop connectivity patterns between the layers in the DDSCAB blocks provide a deep supervision by re-using the feature maps, while dilated layers increase the receptive field exponentially. Spatial attention gates are used in the main building blocks, and encourage the network to concentrate on extracting features from the tumor adjacency. The channel attention gates are used in the skip connections between the contracting and expanding paths of the Unet (named ChA2), for filtering irrelevant feature maps to improve the training process. The proposed network DDAUnet does not include ChA1, and this block is only used in some of the experiments during the optimization of the network configuration.
In section \ref{sec:experimental_results} the performance of DDAUnet will be compared with DUnet \cite{yousefi2018esophageal}, dilated dense unet (DDUnet) which is DUnet with dilated convolutional layers in the dense blocks, DDAUnet without ChA2, i.e. DDAUnet-noChA2, DDAUnet with ChA1 and without SpA and ChA2, i.e. DDAUnet-noSpA-plusChA1-noChA2, DDAUnet with ChA1 and without ChA2, i.e. DDAUnet-plusChA1-noChA2.

According to \cite{cordts2016cityscapes}, the incorporation of a stack of convolutional layers with small receptive fields in the first layers rather than few layers with large receptive fields decreases the number of the parameters, increases non-linearity of the network, and consequently makes training of the network easier. These layers aid the network to extract significant features before applying convolutional operations with a wider receptive field in DDSCAB. Therefore, the network starts with two consecutive 
${(3\times3\times3)\text{conv}_{d=1, p=true} + \text{BN} + \text{ReLU}}$, in which $3\times3\times3$, $d$ and $p$ indicate the kernel size, dilation factor and padding of the convolutional layer respectively. Also, BN and ReLU denote batch normalization and a Rectified linear unit layer, respectively. 

Afterward, the network is followed by a DDSCAB composed of a dilated dense block (DDB) and spatial and channel attention gates. For each  DDB, $R$ is the number of sub-DDBs. In each sub-DDB, there are $R$ number of ${(3\times3\times3)\text{conv}_{d=2, p=true} + \text{BN} + \text{ReLU}}$ and $R$ number of ${(1\times1\times1)\text{conv}_{d=1, p=true} + \text{BN} + \text{ReLU}}$ layers. The output of a DDB is the concatenation of all preceding sub-DDBs. In our prior work, it has been shown that the loop connectivity patterns in dense blocks assist the network to perform better \cite{yousefi2018esophageal}. In DDBs, $(1\times1\times1)\text{conv}$ layers are used as bottleneck layers, which compress the number of feature maps and thus improve computational efficiency \cite{huang2017densely}.
In this paper, the feature maps in each DDB are compressed by a compression coefficient of $\theta$. The output of DDB then is fed to spatial and channel attention gates in order to selectively filter the GTV irrelevant spatial features and feature maps respectively, which leads to improving the training process. 
In the down-sampling path, the DDSCABs are followed by ${(1\times1\times1)\text{conv}_{d=1, p=true} + \text{BN} + \text{ReLU}}$. Using $1\times1\times1$ convolutional layers does not affect the receptive field of the network, however, increases the non-linearity in between layers \cite{simonyan2014very}. 
At the end of down-sampling path and in the up-sampling path every DDSCAB is followed by ${(3\times3\times3)\text{conv}_{d=1}+\text{BN}+\text{ReLU}}$. 
In Section \ref{sec:experimental_results} we will investigate the effect of deploying spatial and channel gates in DDSCAB and will see that utilizing only the spatial gate is more effective. 
Also, the skip connections between the contracting and expanding path are equipped by channel attention gates to filter the irrelevant feature maps. Later we will show that leveraging the spatial and channel attention gates aid the network to end up with better segmentation results. 
The network is finalized by a convolutional layer with linear activation and a soft-max layer to compute a probabilistic output. The probabilistic output can be classified as tumor and non-tumor regions. The skip connections between the down-sampling and up-sampling paths demonstrate cropped concatenation of the feature maps of the corresponding down-sampling levels and up-sampling levels. 

\subsection{Loss function} \label{sec:loss_func}

In this work, similar to our prior work \cite{yousefi2018esophageal} we used the Dice coefficient as our main loss function \cite{milletari2016v}:
\begin{equation}
    \mathrm{DSC_{GTV}}=\frac{2\sum_i^N s_i g_i}{\sum_i^N s_i^2+\sum_i^N g_i^2},\label{eq:DSC}
\end{equation}
where $s_i\in S$ is the binary segmentation of the GTV predicted by the network and $g_i\in G$ is the ground truth segmentation. 
We investigated different combinations of loss functions including boundary loss \cite{kervadec2019boundary}, distance map loss \cite{caliva2019distance}, and focal Dice \cite{wang2018focal}. In \cite{kervadec2019boundary} it is shown that the boundary loss can be approximated by:
\begin{equation}
    {\mathzapf{L}}_B(\theta)= \int_\omega \phi_G(q) s_\theta(q)dq,
\end{equation}
where $\phi_G$ and $s_\theta$ denote the level set representation of the boundary of ground truth, and network output, respectively. 
In Section \ref{sec:experimental_results} it will be discussed that the combination of Dice and boundary loss works the best for this problem.

\section{Data, training details and evaluation}\label{sec:experimental_results}

\subsection{Dataset}
All patients of this study received one of the following two treatments:
\begin{enumerate}
    \item Neoadjuvant chemo-radiotherapy (CRT) followed by
surgical resection. The radiotherapy
is 23 $\times$ 1.8 Gy, 5 fractions a week. The external beam radiotherapy consisted of 23 fractions of 1.8 Gy, five fractions per week. Concurrent
chemotherapy consisted of 5 weekly administrations
of carboplatin and paclitaxel.
    \item If patients are inoperable (proximal tumors they receive primary/definitive chemoradiotherapy (CRT).
The external beam radiotherapy consisted of 28 fractions of 1.8 Gy, five fractions per week. Concurrent chemotherapy consisted of 6 weekly administrations of carboplatin and paclitaxel.
\end{enumerate}

The dataset used in this study consists of 288 distinct patients acquired for a study approved by the Medical Ethics Review Committee of Leiden University Medical Center, the Netherlands.
The dataset includes two sub-datasets from 21 and 267 patients, respectively, in which each patient received either
treatment plan A or B. The data acquisition was performed with a Brilliance Big Bore scanner (Philips Healthcare, Best,Holland) and the delineation process was done by Pinnacle3, (version 9.6–9.8; Philips Radiation Oncology Systems, Fitchburg, WI.) treatment planning software. The ground truth segmentation was performed by the MD on 2D axial slices and evaluated on the 3D cardinal planes.
Table \ref{tab:dataset_details} tabulates the details of the datasets. The first dataset includes five repeat CT scans acquired at different time points. Three time-points contain only one 3D CT scan, and two time-points include one 3D CT scan and one 4D CT scan, composed of 10 breathing phases. Each subdataset includes a corresponding esophageal GTV segmentation for each CT scan, which has been delineated by one (dataset \rom{1}) or multiple (dataset \rom{2}) experienced physicians. Each scan contains 58-108 slices with an image resolution of $512\times512$ pixels and an average voxel thickness of $0.98 \times 0.98 \times 3 \text{mm}^3$, and were re-sampled to a voxel size of $1 \times 1 \times 3 \text{mm}^3$ in this study.

\begin{table*}[]
    \centering
    \caption{Details of the dataset}
    \begin{tabular}{c|c|c|c|c|c}
    \hline 
        dataset & $\#$ of patients& $\#$ of scans &Type&Time period & Treatment plan
        \\\hline
        \multirow{3}{*}{\rom{1}}&\multirow{3}{*}{21 (A)}&\multirow{3}{*}{525}&
        5 time-points
         & \multirow{3}{*}{2012-2014}& \multirow{3}{*}{A: Neoadj. CRT} \\
        && & 2 time-points: 3D scans&&\\
        && & 3 time-points: 3D + 4D scans&&\\\hline
        \multirow{2}{*}{\rom{2}} & \multirow{2}{*}{162 (A) + 105 (B)}& \multirow{2}{*}{267}& \multirow{2}{*}{3D} &\multirow{2}{*}{2014-2019} & A: Neoadj. CRT  \\
        &&&&&B: def. CRT \\
        
    \hline 
    \end{tabular} \label{tab:dataset_details}
\end{table*}

\subsection{Training details}\label{sec:traininf_details}
In this work, the proposed network, which contains 65k trainable parameters, has been implemented in Google’s Tensorflow and experiments are carried out using a NVIDIA Quadro RTX6000 with 24 GB of GPU memory. For all networks, the patch extraction process has been implemented by multi-threaded programming in which fetching the images into RAM, extracting the patches from the fetched images and feeding the extracted patches to the GPU are done concurrently. The multi-threading technique speeds up the patch extraction process. The input patches have been augmented by white noise extracted from a Gaussian distribution of $N(\mu^\prime, \sigma^\prime)$, in which $\mu^\prime = 0$ is the mean of the distribution and $\sigma^\prime$ is the standard deviation, which is selected randomly between 0 and 5. During the test process, the fully convolutional nature of the network is used, with zero padding to yield equal output size. For managing the GPU memory with a larger input patch, we use a batch size of seven. 

For designing the best configuration of the network, we perform experiments comparing different architectures and loss functions. The datasets \rom{1} and \rom{2} are divided randomly into three distinct sets detailed in Table \ref{tab:data_division}. The optimization of the network is performed on the validation set. The test set is excluded from the model optimization and kept independently for the final evaluation. After choosing the best configuration of the network, the final model is trained for two more random splits of the training and validation sets, resulting in three trained models. At the end, an average of the final results for the chosen network, trained on three splits, is reported on the test set.
\begin{table}[]
    \centering
    \begin{tabular}{l|c|c|c||c}
    \hline
       dataset & $\#$ of P/S& DB \rom{1} & DB \rom{2} & total\\\hline
       \multirow{2}{*}{training}  &P&13&182&195\\
       &S& 325&182&507\\\hline
       \multirow{2}{*}{validation}  &P&2&23&25\\
        &S&50&23&53\\\hline
       \multirow{2}{*}{testing}&P& 6 & 62&68\\
       &S&150&62&212\\
       \hline\hline
       \multirow{2}{*}{total}&P&21&267&288\\
       &S&525&267&792\\
       \hline
    \end{tabular}
    \caption{Data split into training, validation and testing sets. P and S denote distinct patients and scans.}
    \label{tab:data_division}
\end{table}

In Section \ref{sec:results_alanysis} the optimization of the network configuration will be discussed on the validation set. Then the best network is trained by different linear combinations of the loss functions including Dice, boundary loss, distance map, and Focal loss. 
Then in Section \ref{sec:opt_res}, for reproducibility, the results of the best configuration of the network will be reported for two more distinct and random splits. 

\subsection{Evaluation measures}
For evaluating the results we report $\mathrm{DSC}$ value (see Section \ref{sec:loss_func}), $\mathrm{MSD}$ and Hausdorff distance $(\mathrm{HD})$ which are defined as:
\begin{align}
    \mathrm{MSD} &= \frac{1}{2}\left(\frac{1}{n} \sum_{i=1}^n d(a_i,S) +\frac{1}{m} \sum_{i=1}^m d(b_i,S)\right), \\
    \mathrm{HD} &= \max \{{\max}_i\{d(a_i,S)\},{\max}_j\{d(b_i,G)\}\},
\end{align}
in which $S$ and $G$ are the predicted and ground truth contours, and $\{a_1,...,a_n\}$ and $\{b_1,...,b_m\}$ the surface mesh points of $S$ and $G$, and $d(a_i,S)=min_j\norm {b_j-a_i}$ respectively. For the Hausdorff distance we report the 95\% percentile instead of the maximum for robustness against outliers.
Since defining the slices where the tumor starts and stops is difficult even for medical doctors, we report perpendicular cranial and caudal distance between the output of the CNNs and the ground truth. The cranial distance (CrD) error is computed as the topmost slice number of the ground truth minus the topmost slice number of the CNN prediction; the caudal distance (CaD) error is computed similarly.

\section{Experimental results}\label{sec:results_alanysis}

In this section the experimental results are reported, with the datasets divided into training, validation and test sets as described in Section \ref{sec:traininf_details}. Model optimization experiments are described in Section \ref{sec:model_op}, where comparison is performed on the validation set. Subsequently, the final results are reported on the test set in Section \ref{sec:opt_res}. For all experiments, we extract the largest component of the network prediction using connected component analysis, and report that. A repeated measure oneway ANOVA test was performed on the Dice values using a significance level of p = 0.05.

\subsection{Model optimization}\label{sec:model_op}

\begin{figure*}[ht]
  \centering
  {\includegraphics[width=18cm,clip]{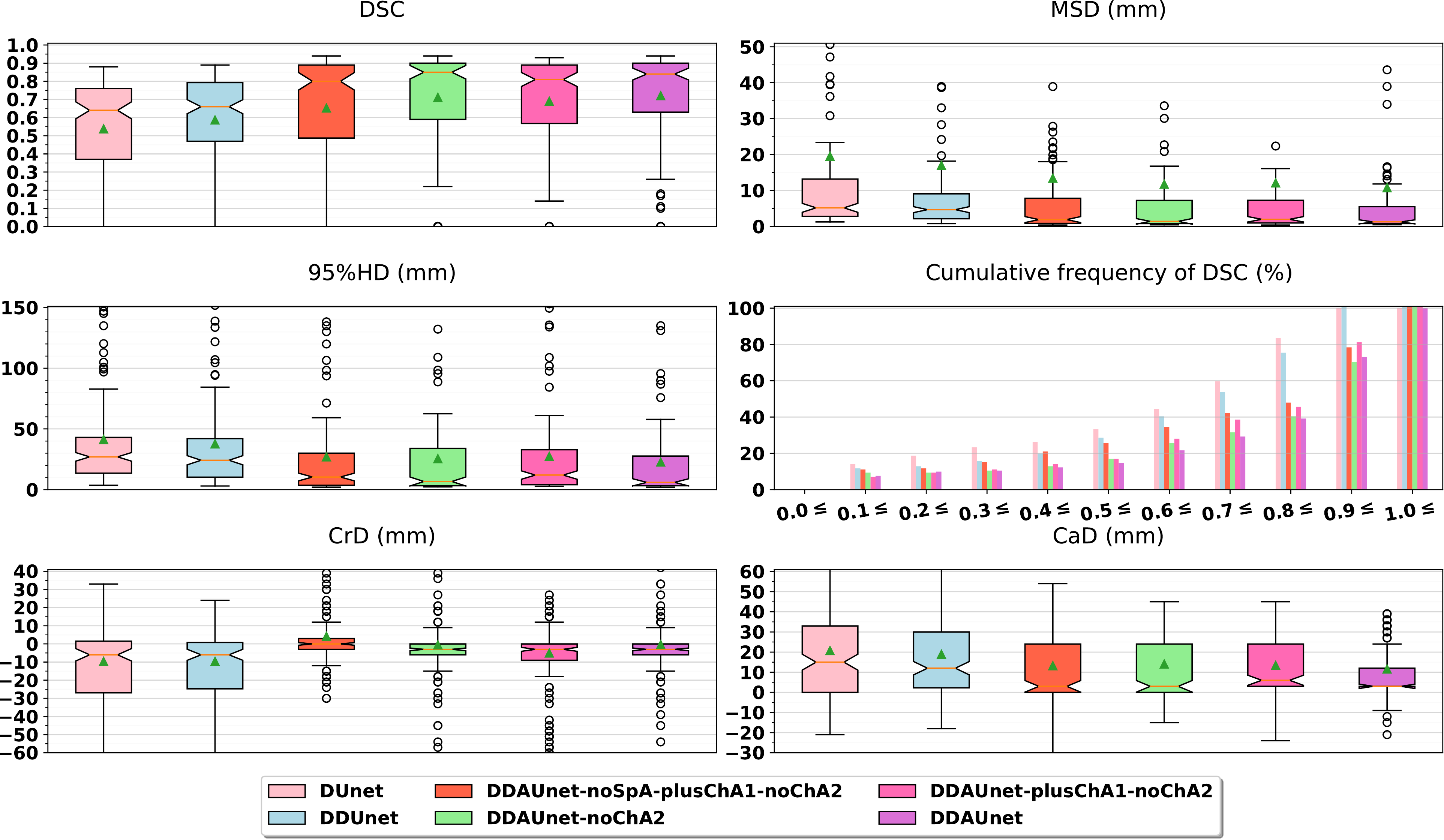}}
  \put (-478,284) {\footnotesize *}\put (-478,190) {\footnotesize 7}  \put (-476,98) {\footnotesize5}
  \put (-222,284) {\footnotesize17}  \put (-220,98) {\footnotesize6}
  \put (-438,284) {\footnotesize *}\put (-438,190) {\footnotesize7}  \put (-438,98) {\footnotesize9}
  \put (-182,284) {\footnotesize17}  \put (-180,98) {\footnotesize7}
  \put (-398,284) {\footnotesize *}\put (-400,190) {\footnotesize4}  \put (-400,98) {\footnotesize15}
  \put (-142,284) {\footnotesize20}  \put (-140,98) {\footnotesize3}
  \put (-360,190) {\footnotesize11}  \put (-361,98) {\footnotesize16}
  \put (-104,284) {\footnotesize11}  \put (-102,98) {\footnotesize4}
  \put (-320,284) {\footnotesize *}\put (-320,190) {\footnotesize5}  \put (-322,98) {\footnotesize18}
  \put (-64,284) {\footnotesize12}  \put (-62,98) {\footnotesize4}
  \put (-282,190) {\footnotesize5}  \put (-280,98) {\footnotesize1}
  \put (-25,284) {\footnotesize10}  \put (-26,98) {\footnotesize 11}
  \caption{A comparison between different network configurations on the validation set. The number of results with values larger than the maximum value on the vertical axis, is shown on top of each plot. The stars in the DSC plot show statistical significance between DDAUnet and the other CNNs.} \label{fig:boxplot_config} 
\end{figure*}
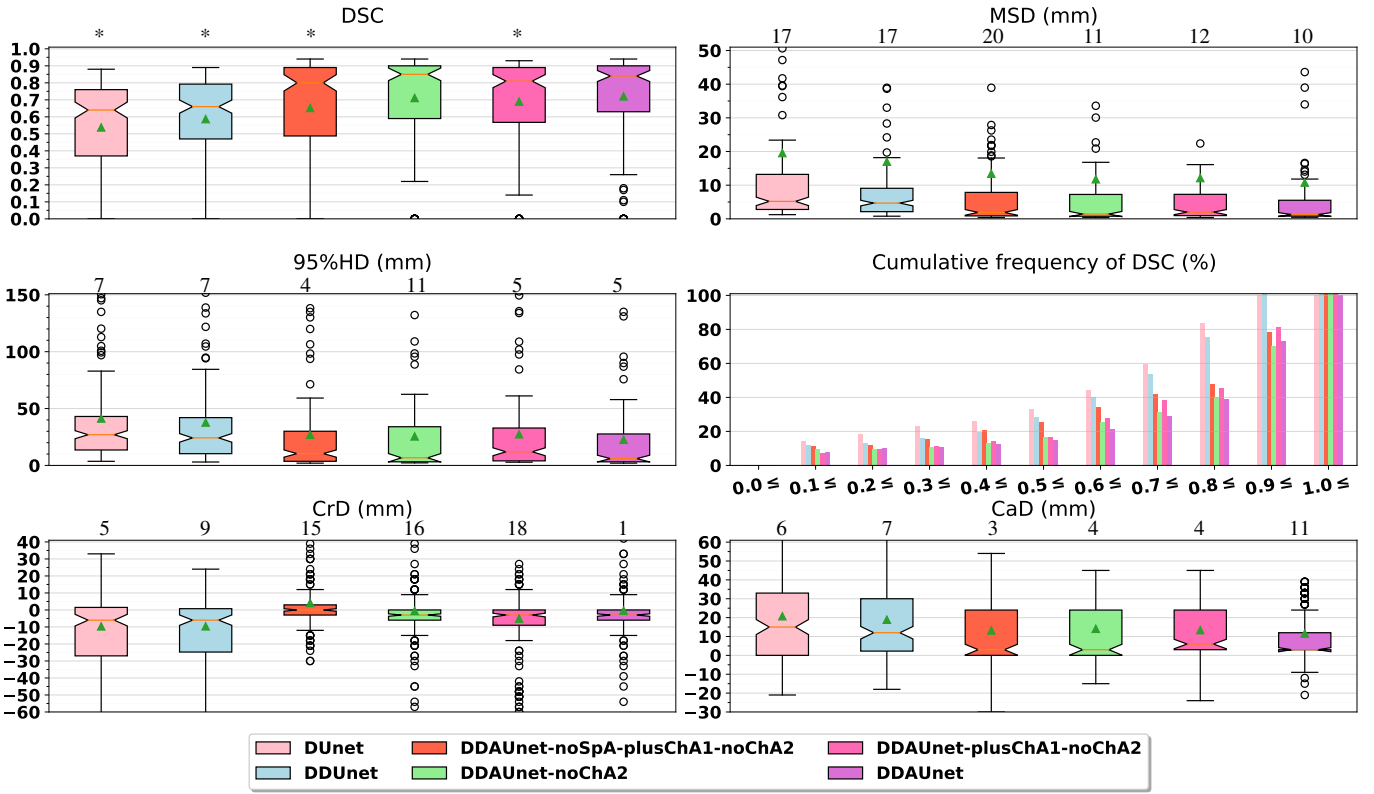

Figure \ref{fig:boxplot_config} shows boxplots of DSC, MSD, 95$\%\mathrm{HD}$, cumulative frequency ($\%$) of $\mathrm{DSC}$, and perpendicular cranial and caudal distance errors for different configurations of the CNN models using the DSC loss function.
Since channel attention gates inside the DDSCAB block, i.e. ChA1 in Fig. \ref{fig:architecture}, did not improve the results, these are not used in the final configuration. The results show that DDAUnet outperforms the other network configurations significantly. 

\begin{figure}[htb]
  \centering
    \includegraphics[width=8cm,trim={0 0cm 0 .0cm},clip]{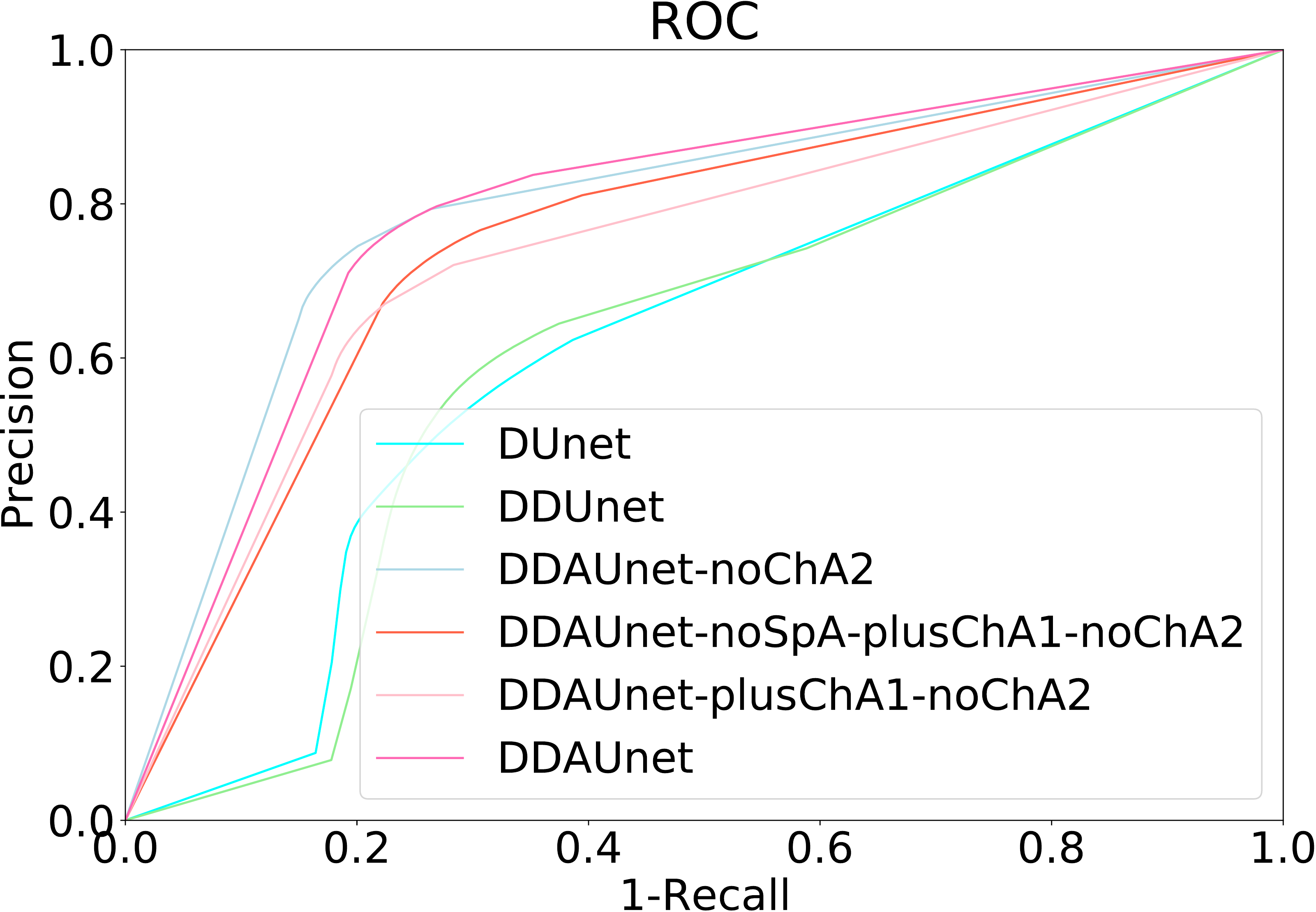}
  \caption{Precision-recall curves for the different network architectures on the validation set using the DSC loss function.}   
  \label{fig:precision-recall}
\end{figure}

Figure \ref{fig:precision-recall} shows the precision-recall curves for the networks on the validation set using the DSC loss function. The precision and recall were calculated with different threshold values applied to the probabilistic output of the networks. For acquiring the final segmentation we used a threshold of 0.5. Table \ref{tab:roc} tabulates the values of the area under the curve (AUC) for the networks on the validation set. The AUC for {DDAUnet} is the largest, and we choose this method as the final network architecture.

\begin{table}[]
    \centering
     \caption{AUC for the networks on the validation set}%
    \begin{tabular}{lc} \hline
      \multicolumn{1}{c}{model} &AUC \\
      \hline
      DUnet &0.49\\
      DDUnet&0.53\\
      DDAUnet-noChA2&0.73\\
      DDAUnet-plusChA1-noChA2&0.63\\
      DDAUnet-noSpA-plusChA1-noChA2&0.71\\
      DDAUnet&0.76\\
        \hline
      \end{tabular}\label{tab:roc}
\end{table}

\begin{figure*}[htb]
  \centering
    \includegraphics[width=18cm,clip]{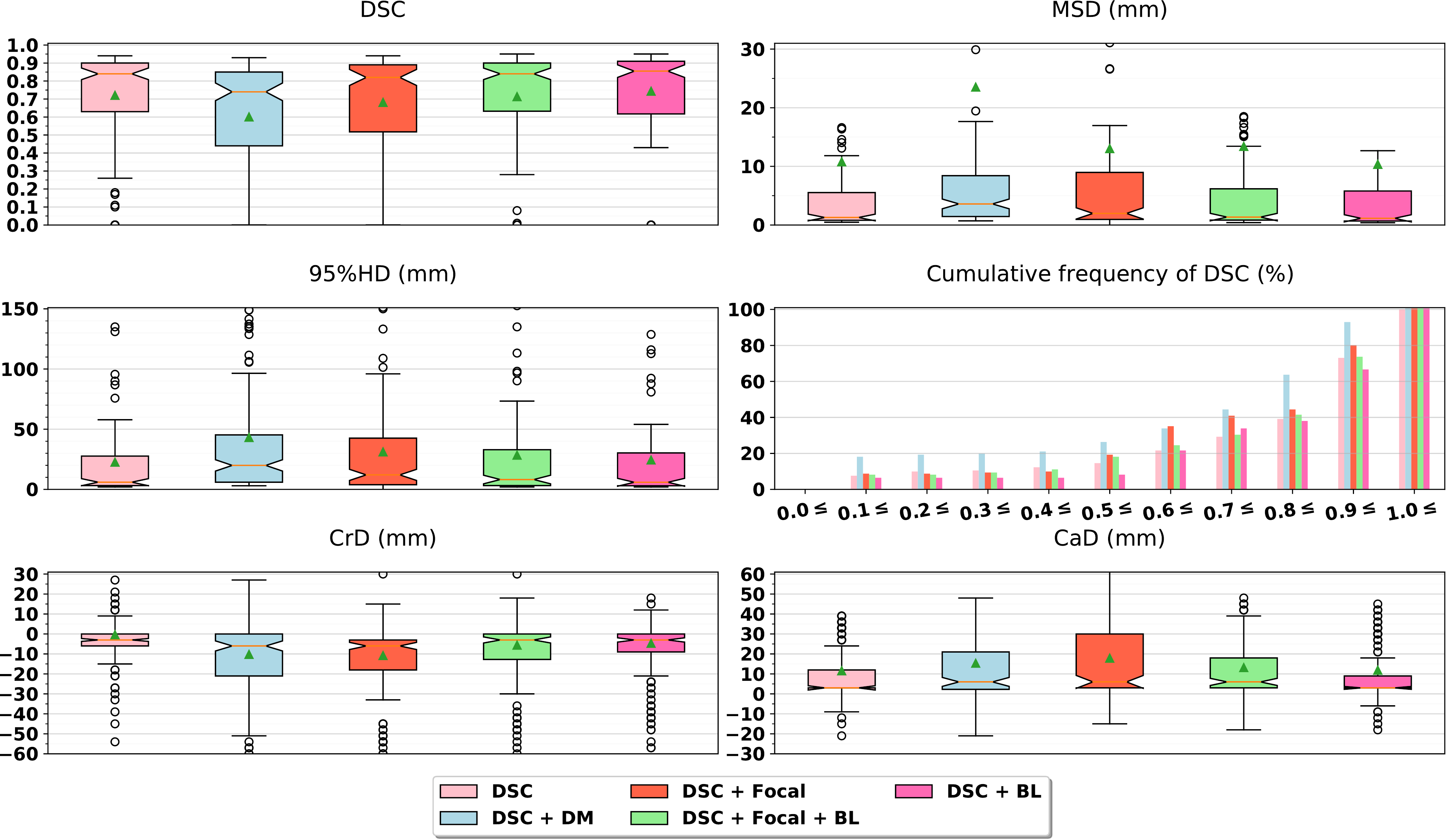}
  \put (-474,284) {\footnotesize*}\put (-474,190) {\footnotesize5}  \put (-474,96) {\footnotesize9}
  \put (-426,284) {\footnotesize*}\put (-218,284) {\footnotesize13}  \put (-217,96) {\footnotesize23}
  \put (-378,284) {\footnotesize*}\put (-428,190) {\footnotesize14}  \put (-430,96) {\footnotesize10}
   \put (-332,284) {\footnotesize*}\put (-170,284) {\footnotesize33}  \put (-170,96) {\footnotesize5}
  \put (-380,190) {\footnotesize5}  \put (-380,96) {\footnotesize26}
  \put (-124,284) {\footnotesize15}  \put (-120,96) {\footnotesize5}
  \put (-332,190) {\footnotesize7}  \put (-334,96) {\footnotesize23}
  \put (-78,284) {\footnotesize13}  \put (-75,96) {\footnotesize6}
  \put (-284,190) {\footnotesize5}  \put (-288,96) {\footnotesize18}
  \put (-30,284) {\footnotesize11}  \put (-30,96) {\footnotesize21}
  \caption{A comparison between deploying different loss functions for DDAUnet on the validation set. The number of results with values larger than the maximum value on the vertical axis, is shown on top of each plot. The
stars in the DSC plot show statistical significance between DSC+BL and the other loss functions.}   
  \label{fig:boxplot_loss} 
\end{figure*}

We experimented with different combinations of loss functions including Dice, boundary loss, distance map loss, Focal Dice on the validation set. Figure \ref{fig:boxplot_loss} shows the results. The results show that DDAUnet using the DSC + BL loss function outperforms the other loss functions significantly.

\subsection{Final results}\label{sec:opt_res}

\begin{figure*}
\begin{subfigure}{0.32\textwidth}
\includegraphics[width=\linewidth,height=4.cm]{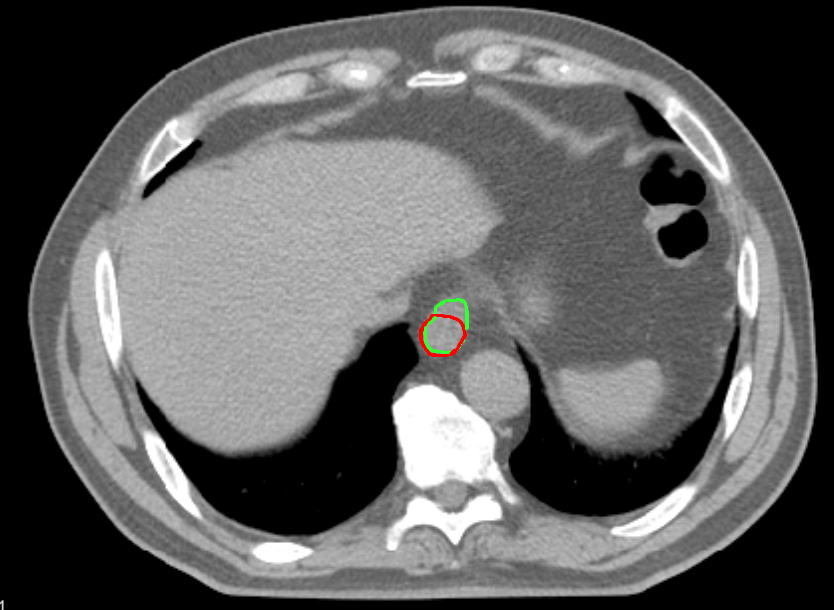}\put (-20,5) {\color{yellow}0.77}

\medskip
\includegraphics[width=0.49\textwidth,height=.1\textheight, trim={8cm 5.cm 6cm 5.5cm}, clip]{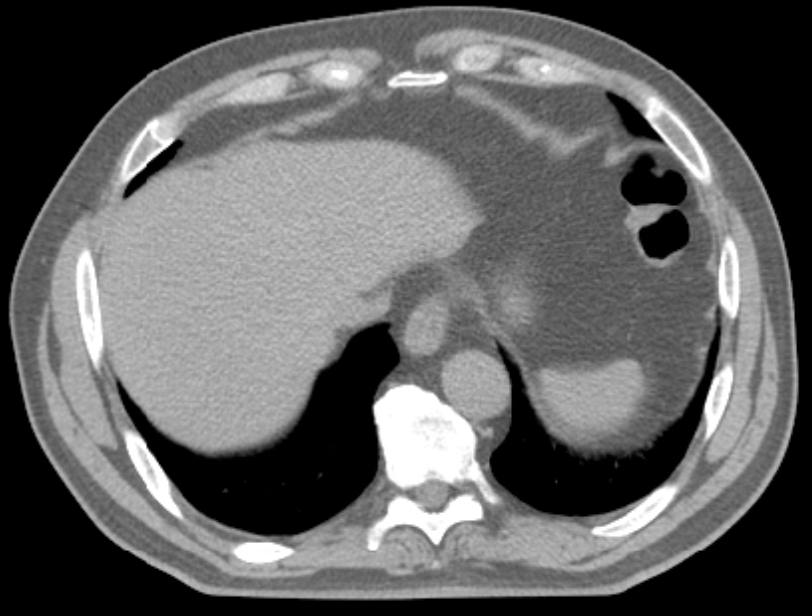}
\hfill
\includegraphics[width=0.49\textwidth,height=.1\textheight, trim={8cm 5.cm 6cm 5.5cm}, clip]{figures/results_final/TEST091_2017-03-13_122_0.7674418603535695.png}
\caption{Normal esophageal GTV}
\end{subfigure}
\hfill 
\begin{subfigure}{0.32\textwidth}
\includegraphics[width=\linewidth,height=4.cm]{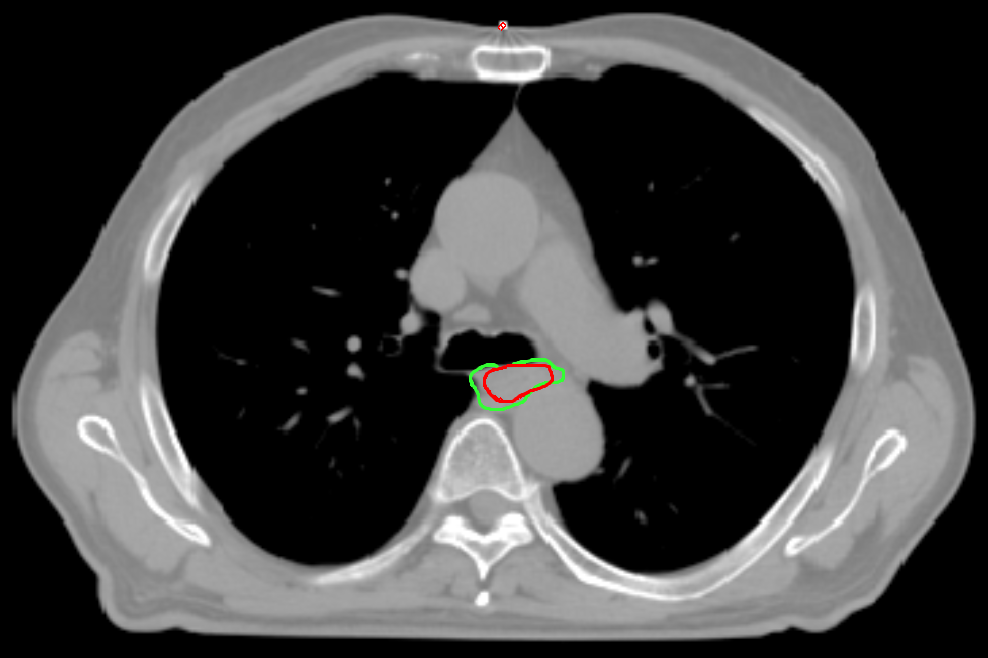}\put (-20,5) {\color{yellow}0.73}

\medskip
\includegraphics[width=0.49\textwidth,height=.1\textheight, trim={9cm 5.cm 8cm 6.5cm}, clip]{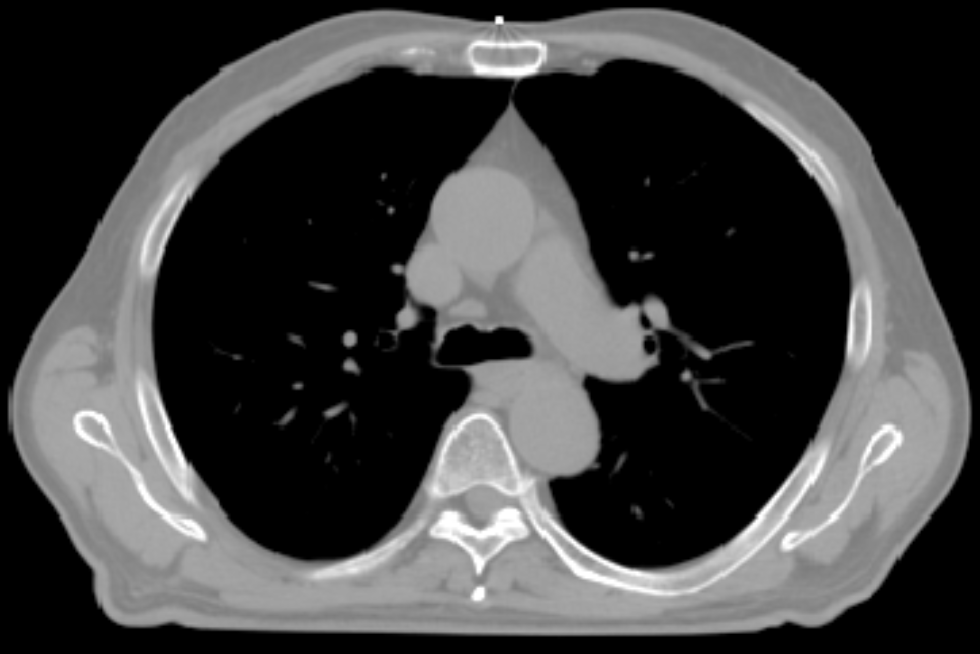}%
\hfill
\includegraphics[width=0.49\textwidth,height=.1\textheight, trim={9cm 5.cm 8cm 6.5cm}, clip]{figures/results_final/zz3710236225_zz3710236225_128_0.7304964537711887.png}
\caption{Normal esophageal GTV}
\end{subfigure}
\hfill 
\begin{subfigure}{0.32\textwidth}
\includegraphics[width=\linewidth,height=4.cm]{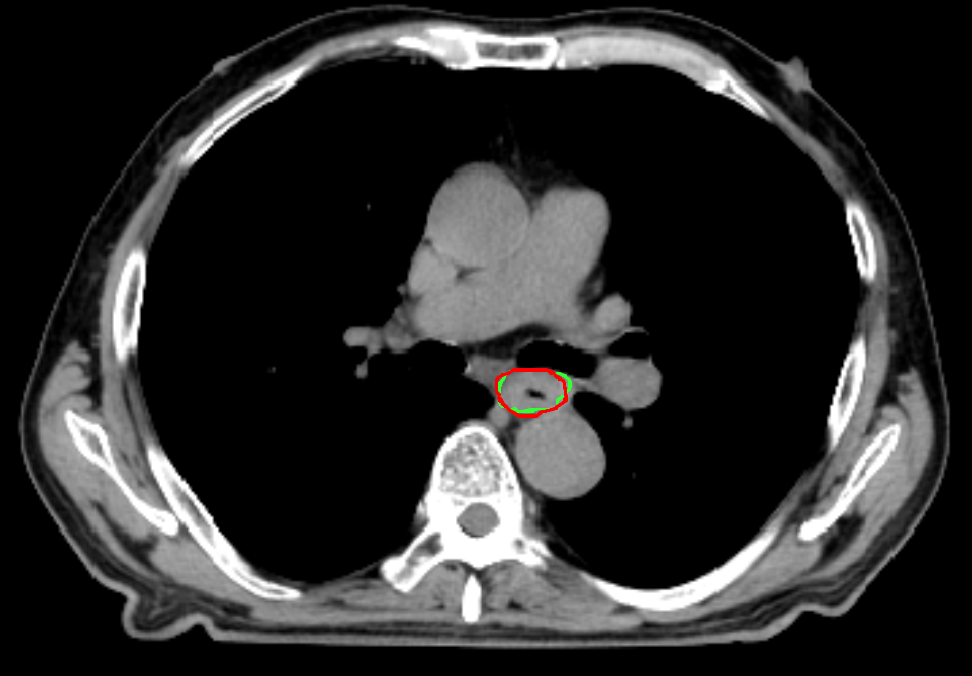}\put (-20,5) {\color{yellow}0.93}

\medskip
\includegraphics[width=0.49\textwidth,height=.1\textheight, trim={10cm 5cm 7cm 6.5cm}, clip]{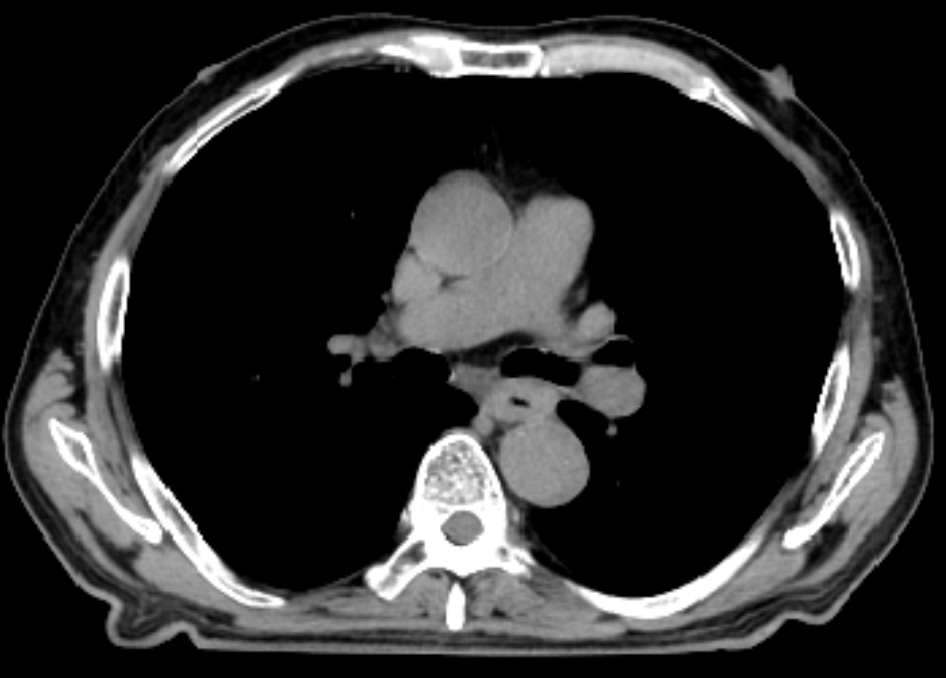}%
\hfill
\includegraphics[width=0.49\textwidth,height=.1\textheight, trim={10cm 5cm 7cm 6.5cm}, clip]{figures/results_final/zz3710236225_zz3710236225_122_0.9258649092379135.png}
\caption{GTV including an \mbox{air pocket}}
\end{subfigure}
\hfill 
\begin{subfigure}{0.32\textwidth}
\includegraphics[width=\linewidth,height=4.3cm]{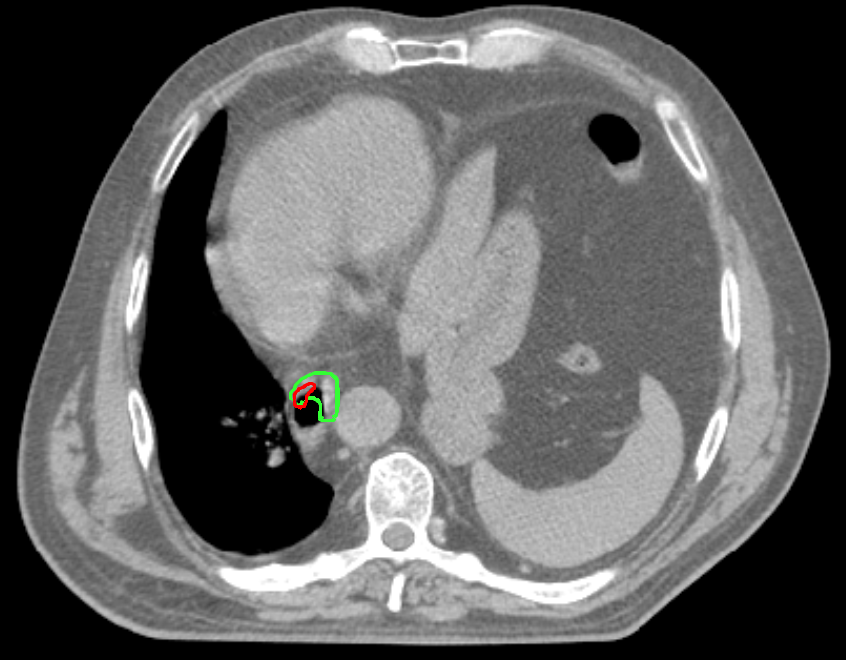}\put (-20,5) {\color{yellow}0.26}

\medskip
\includegraphics[width=0.49\textwidth,height=.1\textheight, trim={5.cm 5.0cm 9cm 7.5cm}, clip]{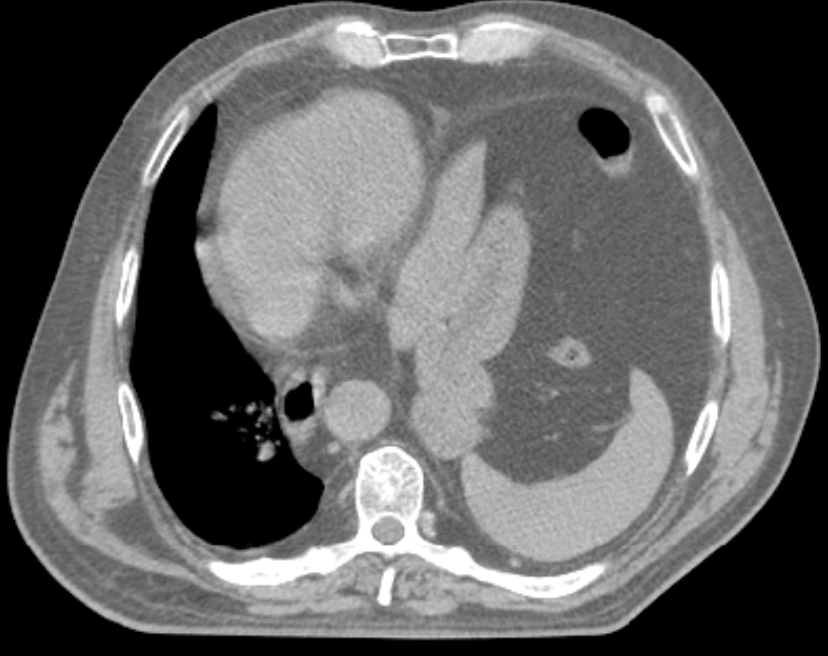}%
\hfill
\includegraphics[width=0.49\textwidth,height=.1\textheight, trim={5.cm 5.0cm 9cm 7.5cm}, clip]{figures/results_final/TEST086_2016-12-21_126_0.25573770483418434.png}
\caption{GTV in a dislocated esophagus  }
\end{subfigure}
\hfill 
\begin{subfigure}{0.32\textwidth}
\includegraphics[width=\linewidth,height=4.3cm]{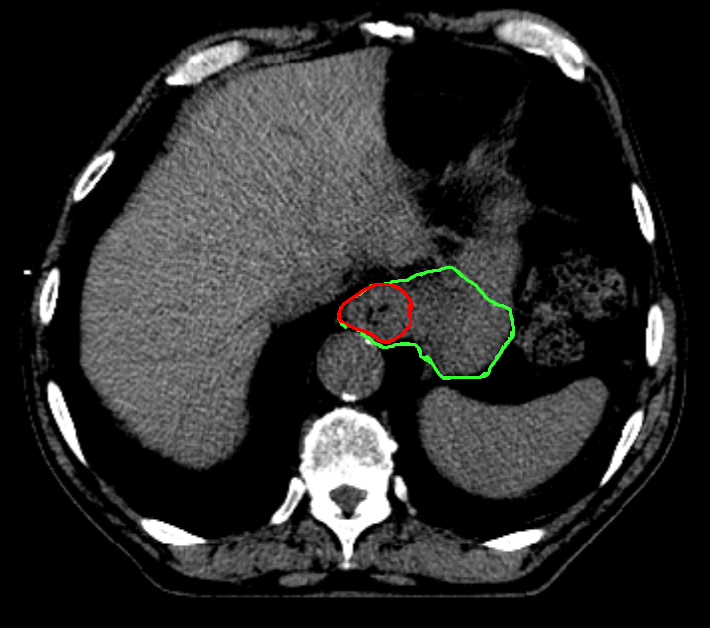}\put (-20,5) {\color{yellow}0.38}

\medskip
\includegraphics[width=0.49\textwidth,height=.1\textheight, trim={6.5cm 4.5cm 4cm 6cm}, clip]{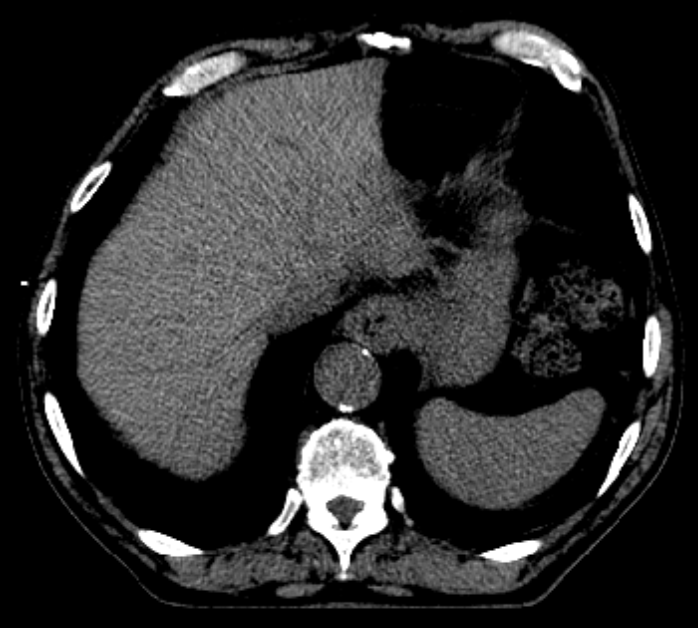}%
\hfill
\includegraphics[width=0.49\textwidth,height=.1\textheight, trim={6.5cm 4.5cm 4cm 6cm}, clip]{figures/results_final/TEST108_2017-09-18_121_0.38301462316133256.png}
\caption{Junction GTV}
\end{subfigure}
\hfill 
\begin{subfigure}{0.32\textwidth}
\includegraphics[width=\linewidth,height=4.3cm]{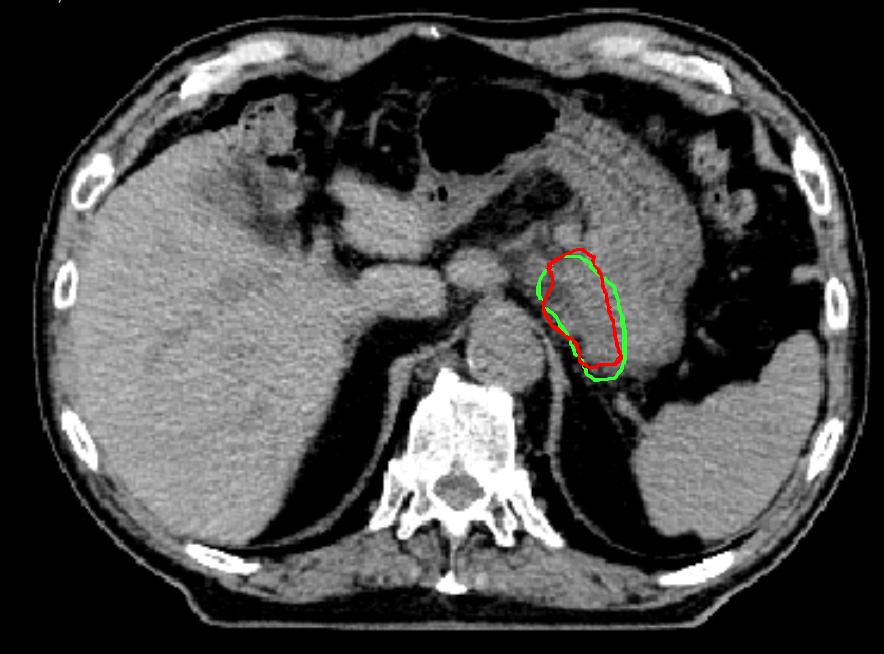}\put (-20,5) {\color{yellow}0.84}

\medskip
\includegraphics[width=0.49\textwidth,height=.1\textheight, trim={7cm 4.cm 4cm 5cm}, clip]{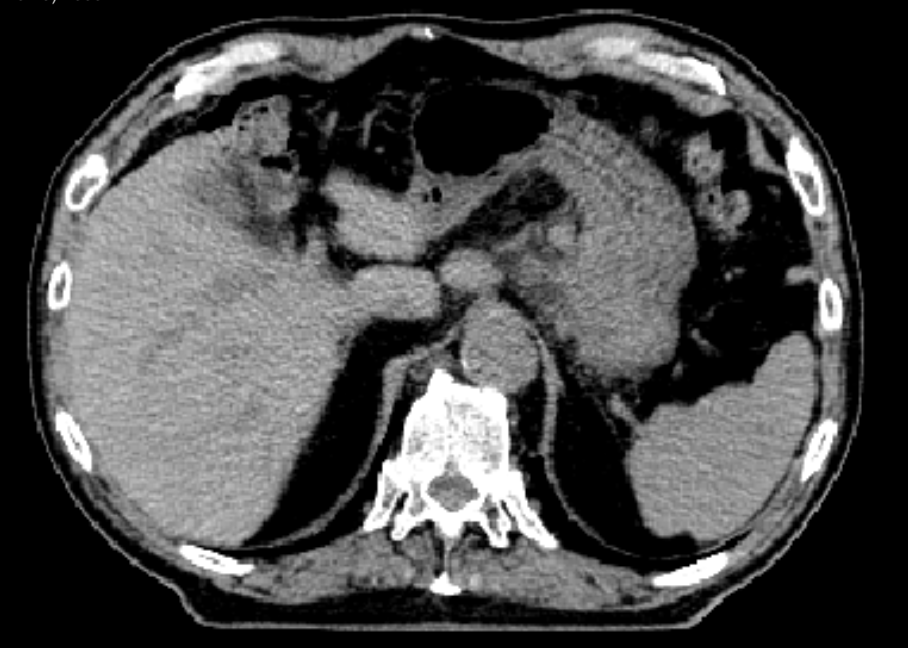}%
\hfill
\includegraphics[width=0.49\textwidth,height=.1\textheight, trim={7cm 4.cm 4cm 5cm}, clip]{figures/results_final/zz3466954516_zz3466954516_113_0.8392504930552638.png}
\caption{Junction GTV}
\end{subfigure}
\hfill 
\begin{subfigure}{0.32\textwidth}
\includegraphics[width=\linewidth,height=3cm]{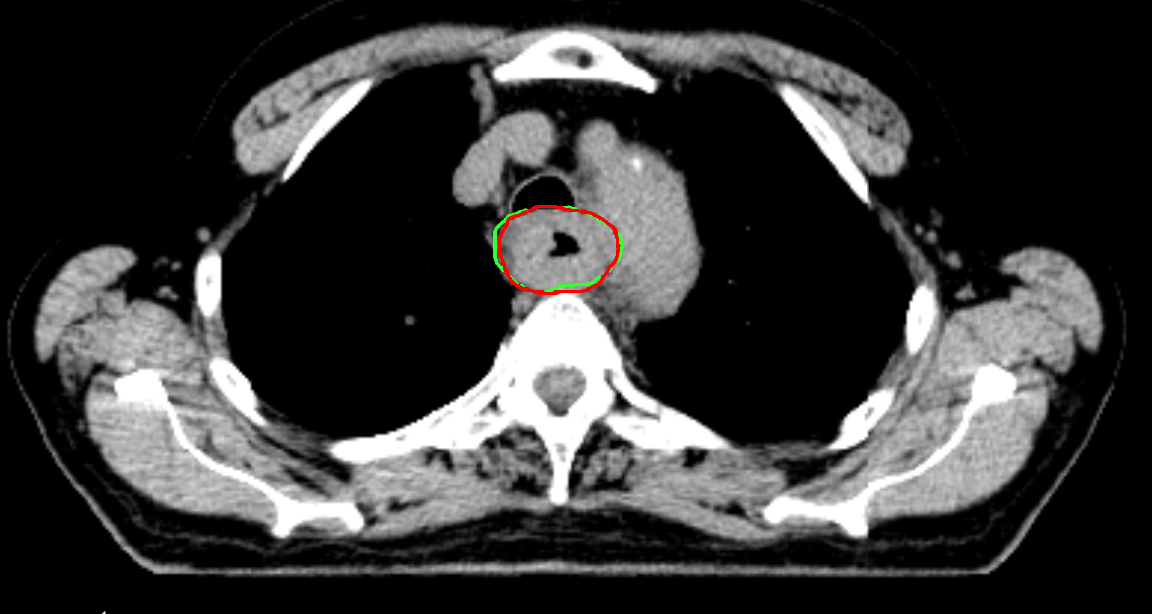}\put (-20,5) {\color{yellow}0.95}

\medskip
\includegraphics[width=0.49\textwidth,height=.1\textheight, trim={10cm 6.cm 10cm 3cm}, clip]{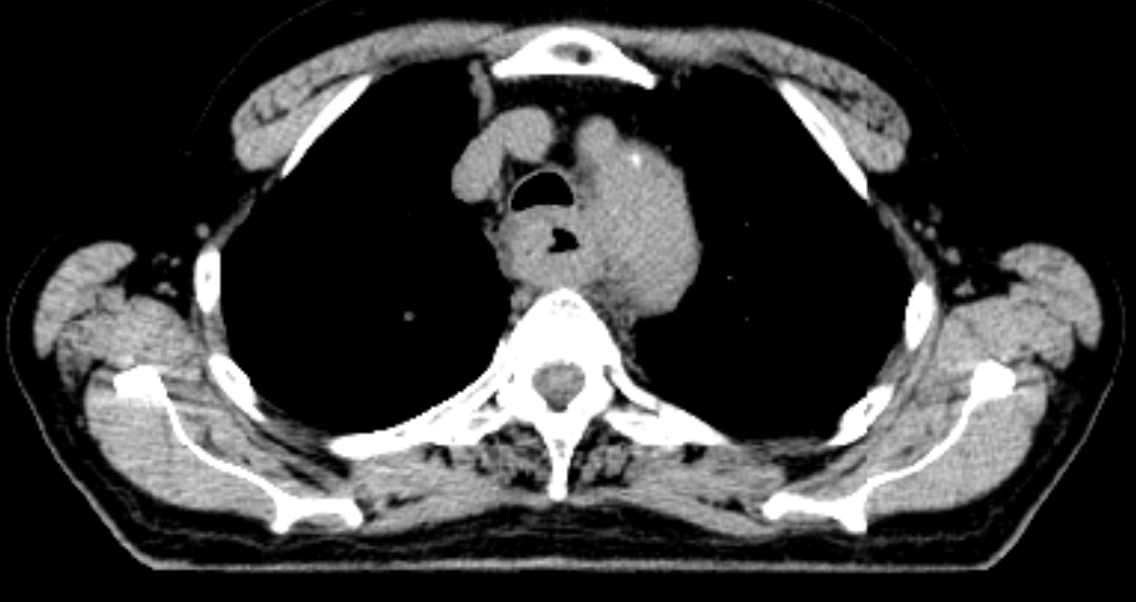}%
\hfill
\includegraphics[width=0.49\textwidth,height=.1\textheight, trim={10cm 6.cm 10cm 3cm}, clip]{figures/results_final/zz197056916_zz197056916_120_0.9516770892011553.png}
\caption{GTV including an \mbox{air pocket}}
\end{subfigure}
\hfill 
\begin{subfigure}{0.32\textwidth}
\includegraphics[width=\linewidth,height=3cm]{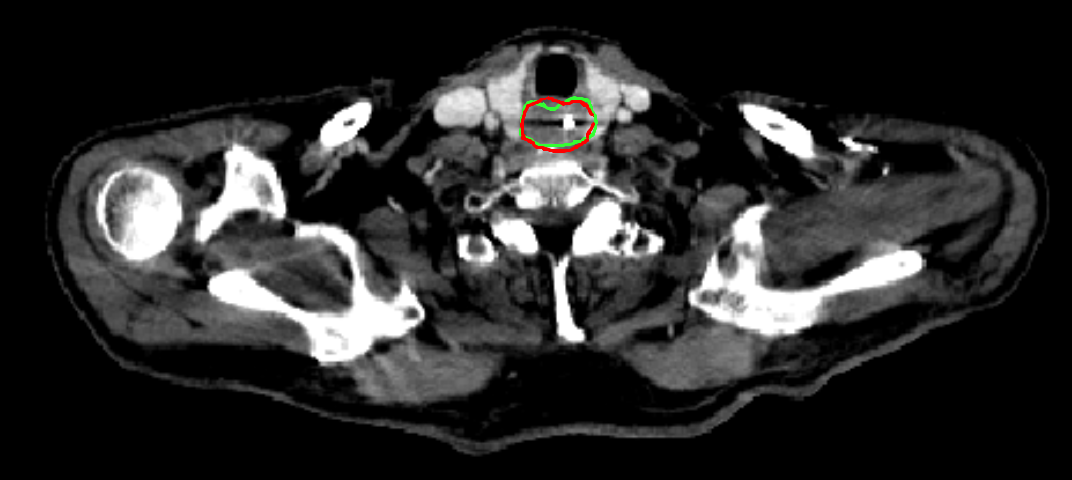}\put (-20,5) {\color{yellow}0.89}

\medskip
\includegraphics[width=0.49\textwidth,height=.1\textheight, trim={10cm 6.cm 10cm 1cm}, clip]{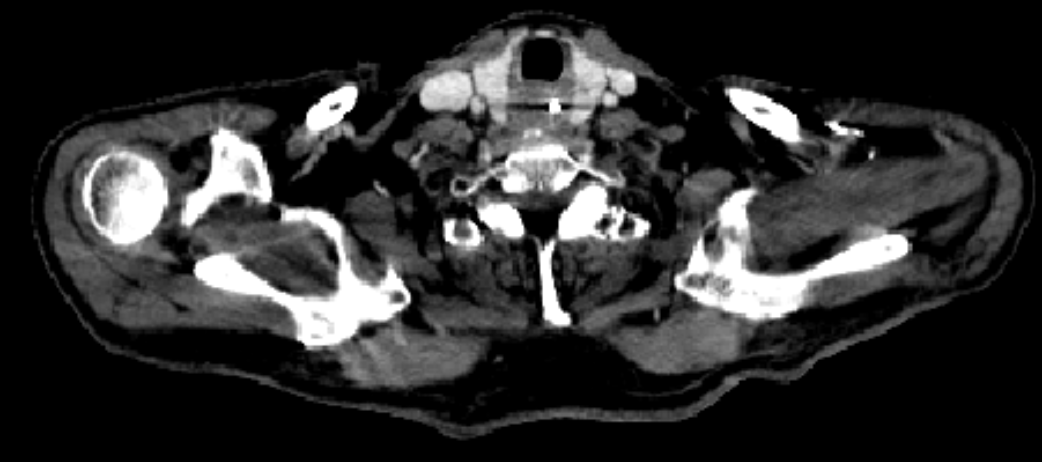}%
\hfill
\includegraphics[width=0.49\textwidth,height=.1\textheight, trim={10cm 6.cm 10cm 1cm}, clip]{figures/results_final/zz3744705519_zz3744705519_121_0.8875502007141014.png}
\caption{Proximal GTV including an \mbox{air pocket} and a feeding tube}
\end{subfigure}
\hfill 
\begin{subfigure}{0.32\textwidth}
\includegraphics[width=\linewidth,height=3cm]{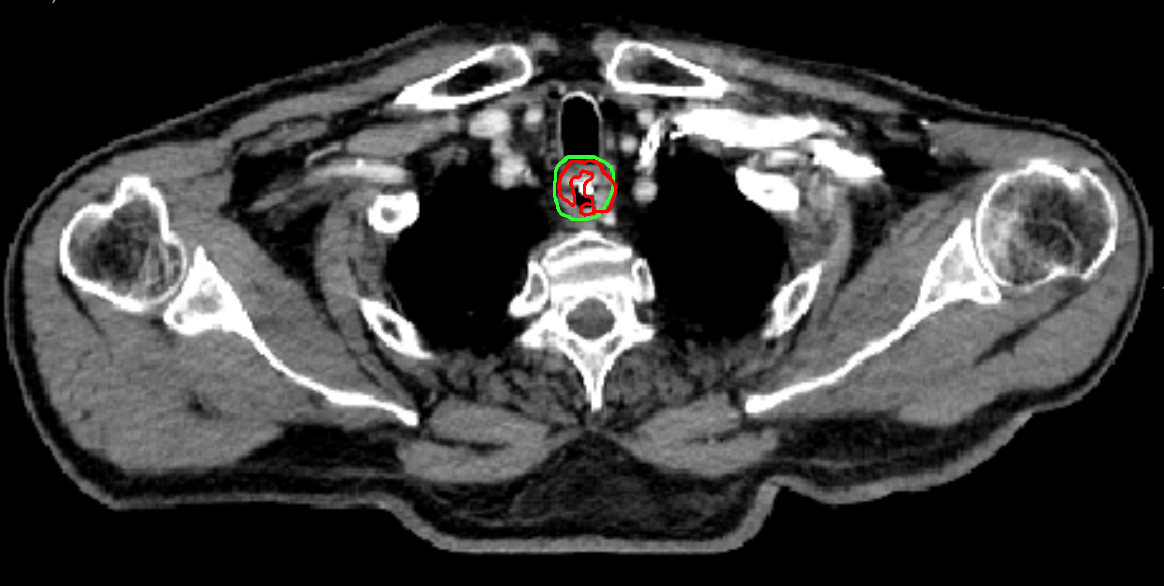}\put (-20,5) {\color{yellow}0.65}

\medskip
\includegraphics[width=0.49\textwidth,height=.1\textheight, trim={12cm 7.cm 11cm 3cm}, clip]{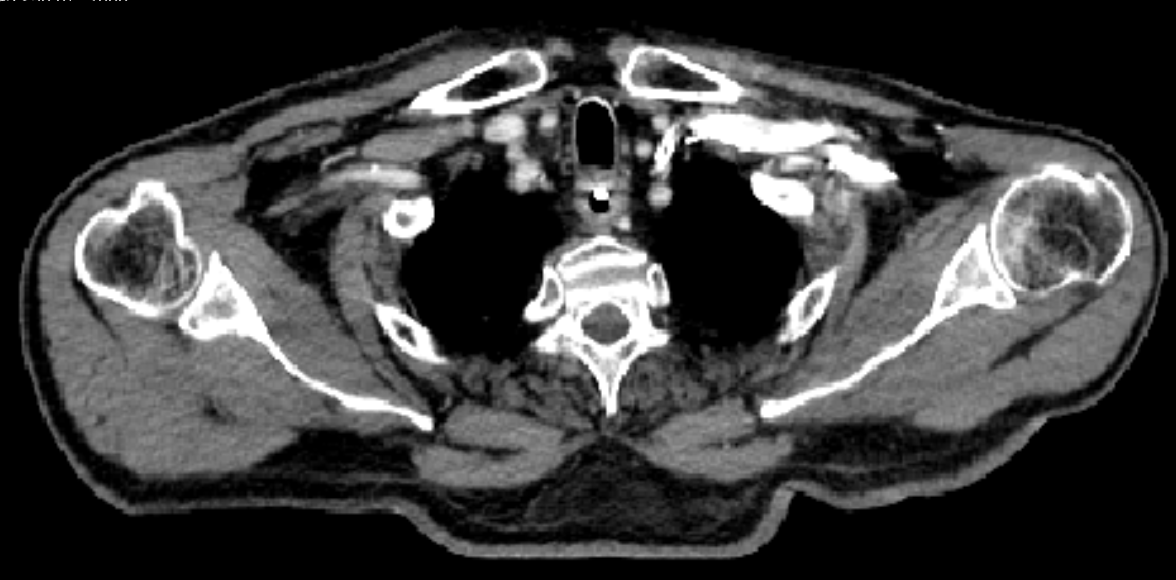}%
\hfill
\includegraphics[width=0.49\textwidth,height=.1\textheight, trim={12cm 7.cm 11cm 3cm}, clip]{figures/results_final/zz3744705519_zz3744705519_111_0.6458333332372272.png}
\caption{Proximal GTV including an \mbox{air pocket} and a feeding tube}
\end{subfigure}
\caption{Example results of the proposed method with 2D $\mathrm{DSC}$ value in yellow. The manual delineation and the network results are shown by green and red contours, respectively.}\label{fig:final_results}
\end{figure*}

\begin{figure*} [htp]
\centering
\footnotesize
\begin{tabular}{cccc}
{\includegraphics[width = 0.19\textwidth,trim={7cm 5.1cm 5cm 4.9cm}, clip]{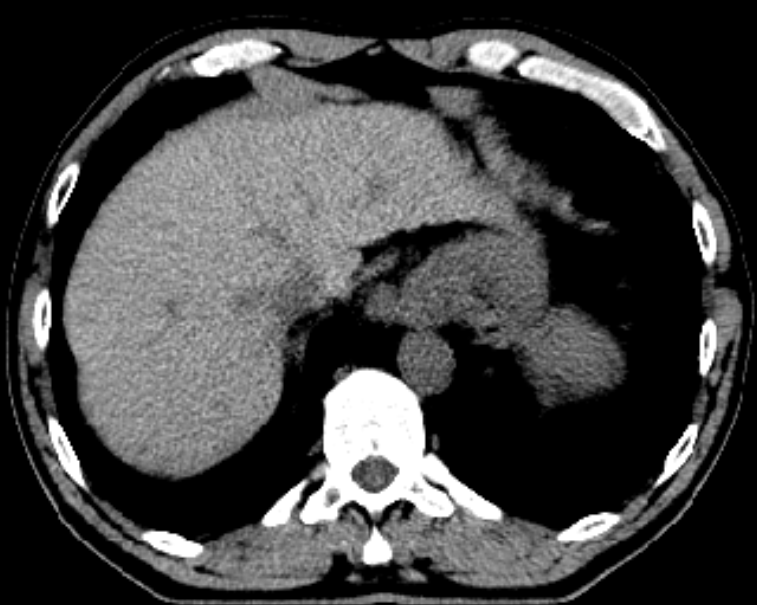}}&
{\includegraphics[width = 0.19\textwidth,trim={7cm 5.cm 5cm 4.8cm}, clip]{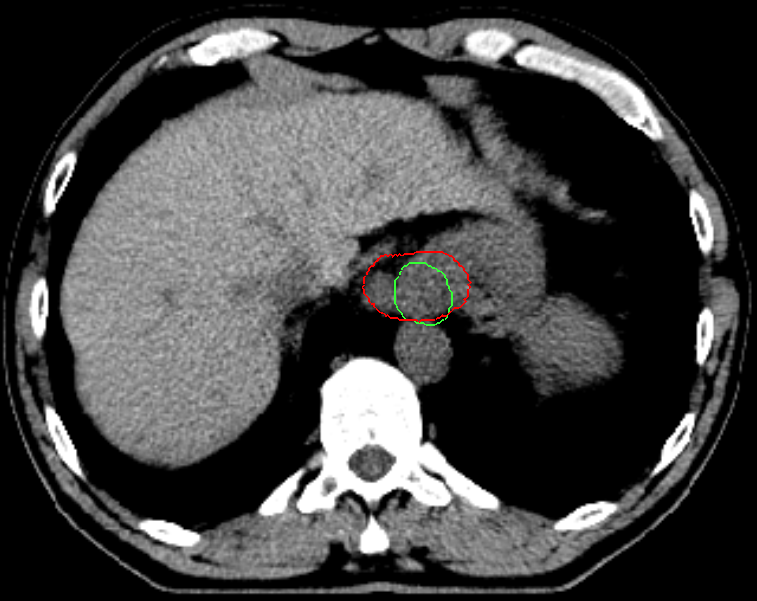}\put (-25,5) {\color{yellow}\normalsize${0.61}$}}&
{\includegraphics[width = 0.19\textwidth,trim={7cm 5.cm 5cm 4.8cm}, clip]{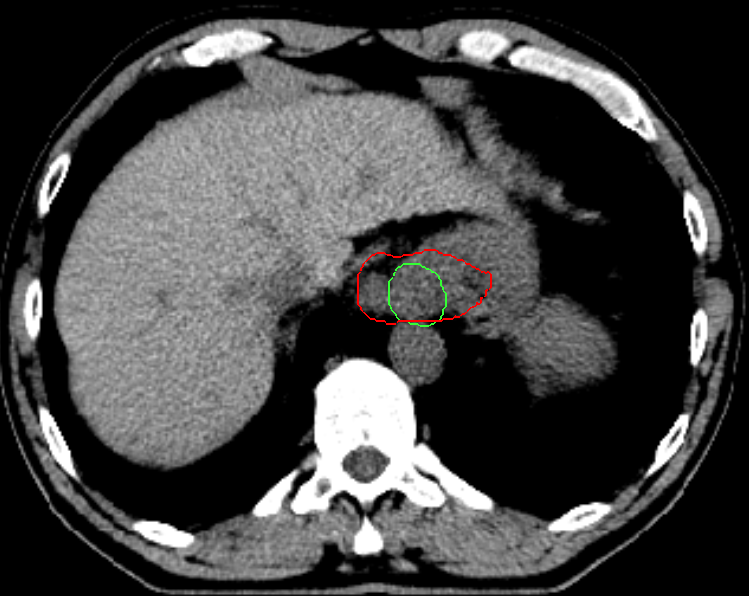}\put (-25,5) {\color{yellow}\normalsize${0.51}$}}&
{\includegraphics[width = 0.19\textwidth,trim={7cm 5.cm 5cm 4.8cm}, clip]{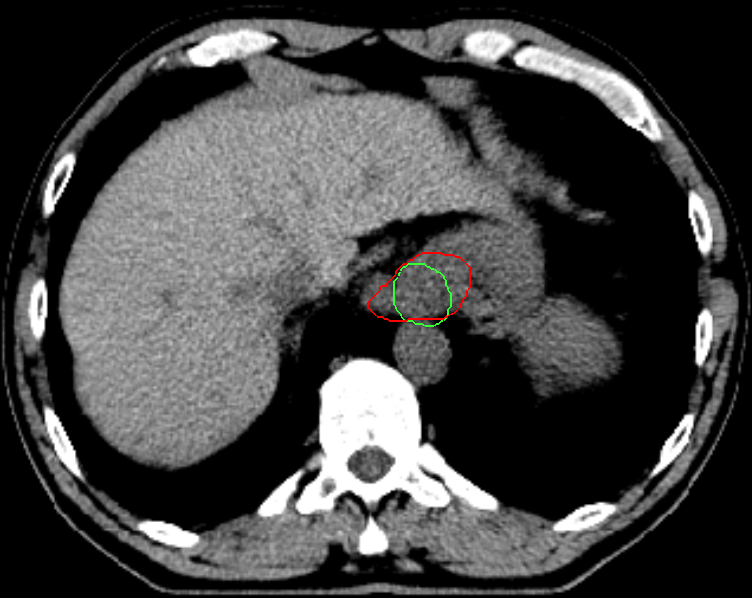}\put (-25,5) {\color{yellow}\normalsize${0.66}$}} \\
{CT scan}  & {DUnet} & {DDUnet}& {DDAUnet-noSpA-plusChA1-noChA2}  \\[0pt]
\end{tabular}
\begin{tabular}{cccc}
{\includegraphics[width = 0.19\textwidth,trim={7cm 5.2cm 5cm 5cm}, clip]{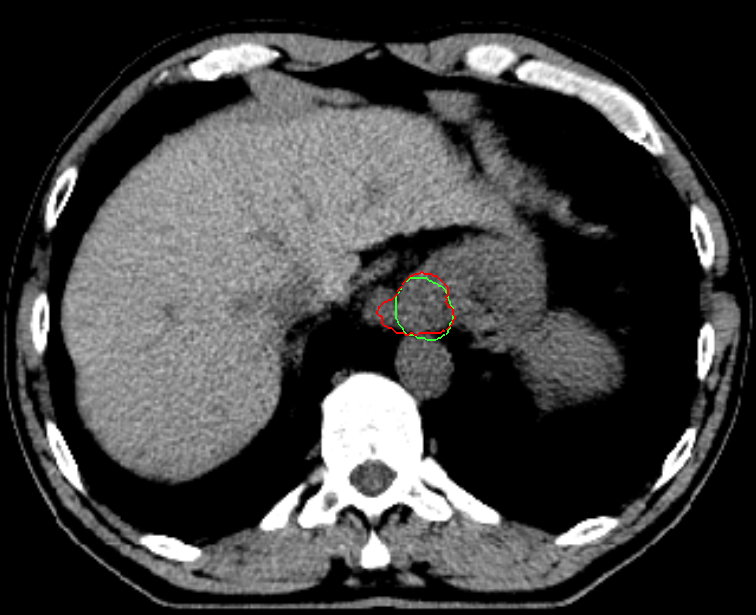}\put (-25,5) {\color{yellow}\normalsize${0.84}$}}&
{\includegraphics[width = 0.19\textwidth,trim={7cm 5.cm 5cm 4.9cm}, clip]{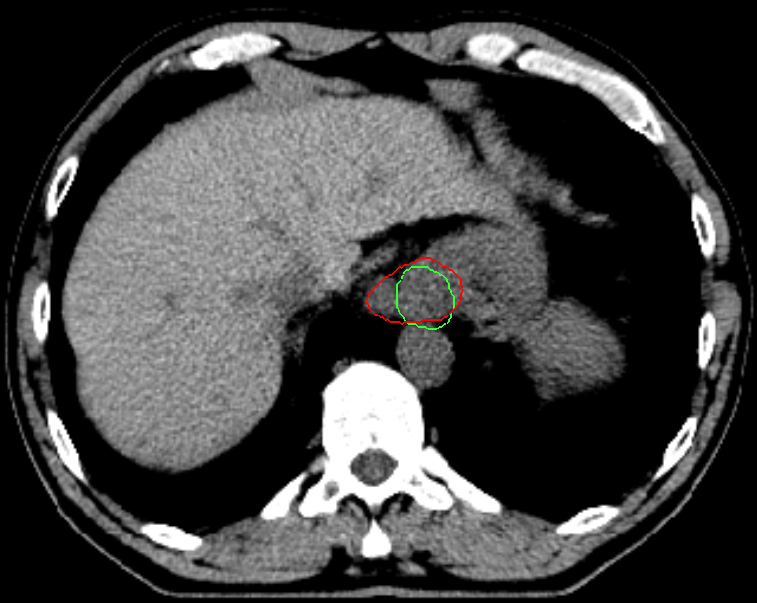}\put (-25,5) {\color{yellow}\normalsize${0.69}$}}&
{\includegraphics[width = 0.19\textwidth,trim={7cm 5.cm 5cm 4.8cm}, clip]{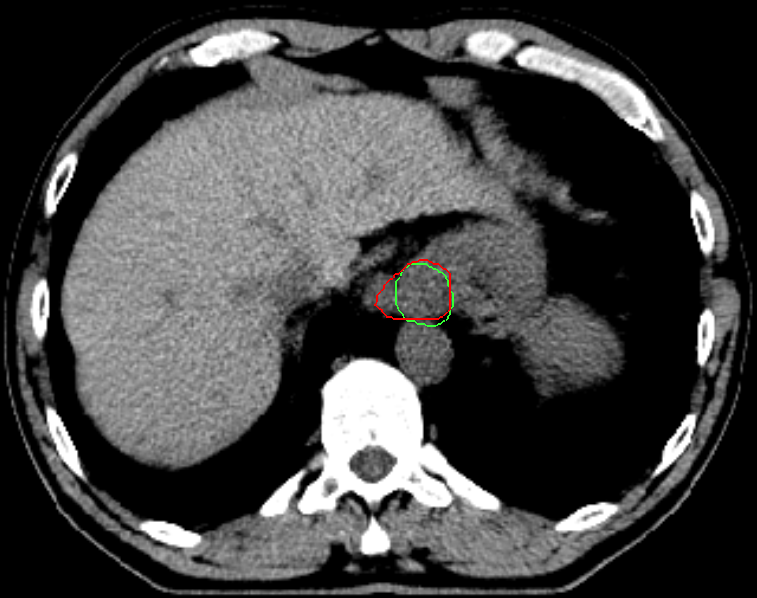}\put (-25,5) {\color{yellow}\normalsize${0.81}$}} \\
{DDAUnet-plusChA1-noChA2}  &{DDAUnet-plusChA1-noChA2}   & {DDAUnet} \\[1pt]
\end{tabular}
\centerline{\color{gray}\rule{13cm}{0.4pt}}
\begin{tabular}{cccc}
&&&\\\includegraphics[width = 0.19\textwidth,trim={7cm 4.cm 5cm 4.8cm}, clip]{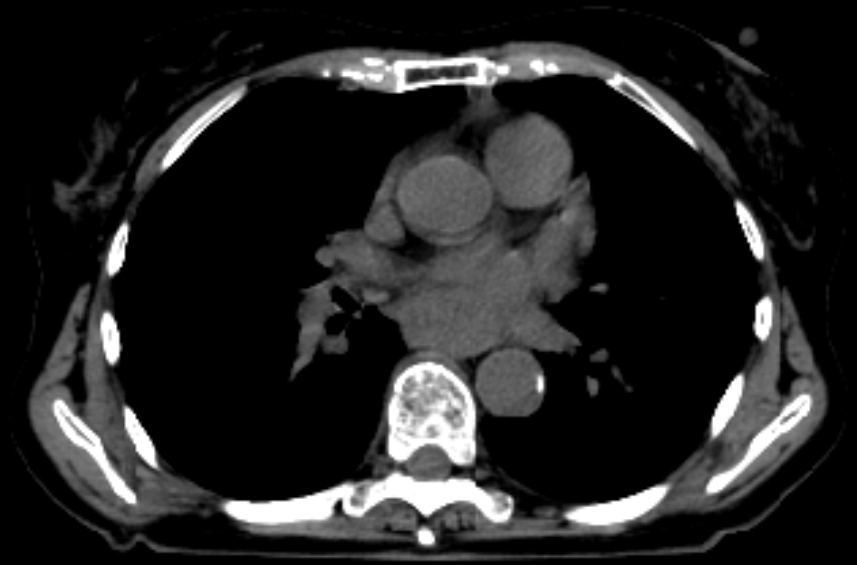}&
{\includegraphics[width = 0.19\textwidth,trim={7cm 4.cm 5cm 4.8cm}, clip]{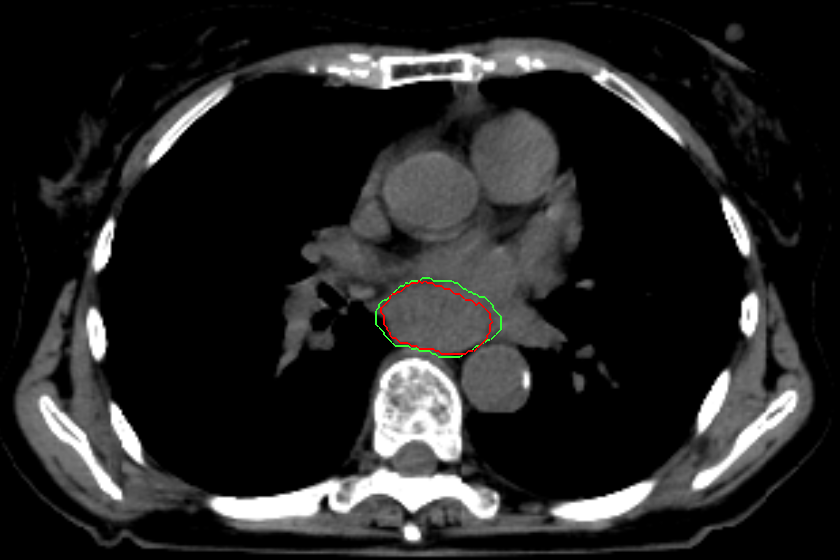}\put (-25,5) {\color{yellow}\normalsize${0.89}$}}&
{\includegraphics[width = 0.19\textwidth,trim={7cm 4.cm 5cm 4.8cm}, clip]{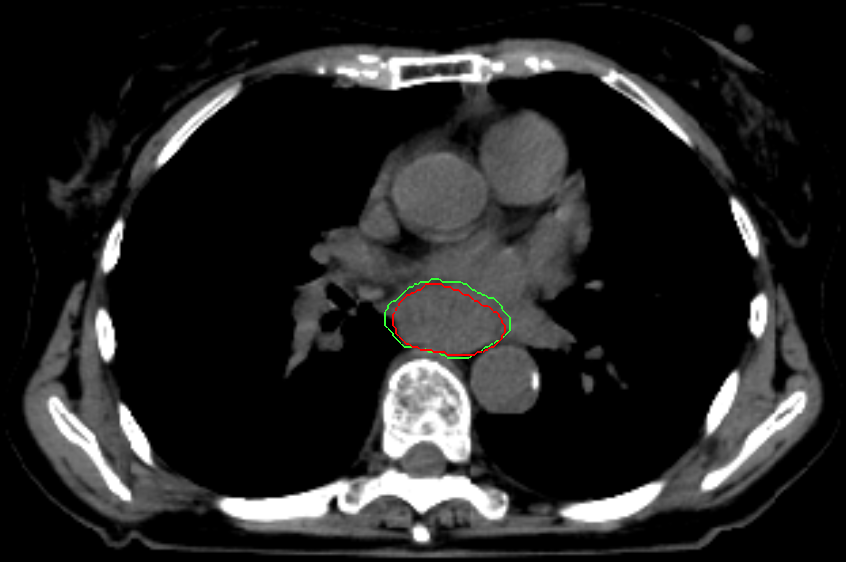}\put (-25,5) {\color{yellow}\normalsize${0.89}$}}&
{\includegraphics[width = 0.19\textwidth,trim={7cm 4.cm 5cm 4.8cm}, clip]{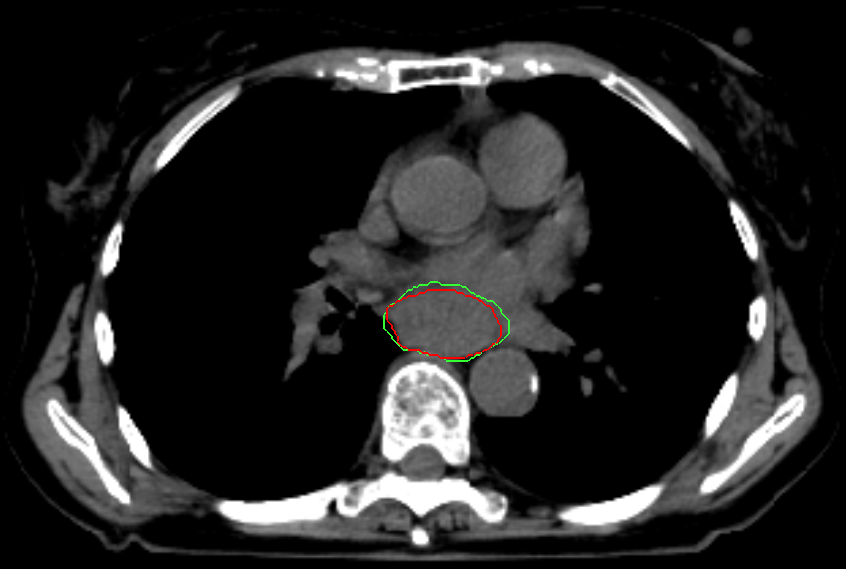}\put (-25,5) {\color{yellow}\normalsize${0.91}$}} \\
{CT scan}  & {DUnet} & {DDUnet}& {DDAUnet-noSpA-plusChA1-noChA2}  \\[0pt]
\end{tabular}
\begin{tabular}{cccc}
{\includegraphics[width = 0.19\textwidth,trim={7cm 4.2cm 5cm 5cm}, clip]{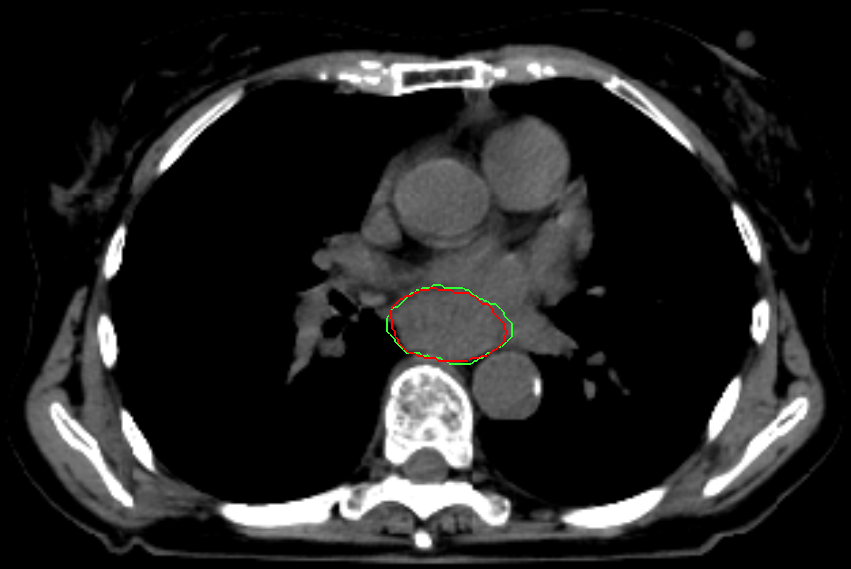}\put (-25,5) {\color{yellow}\normalsize${0.93}$}}&
{\includegraphics[width = 0.19\textwidth,trim={7cm 4.cm 5cm 4.9cm}, clip]{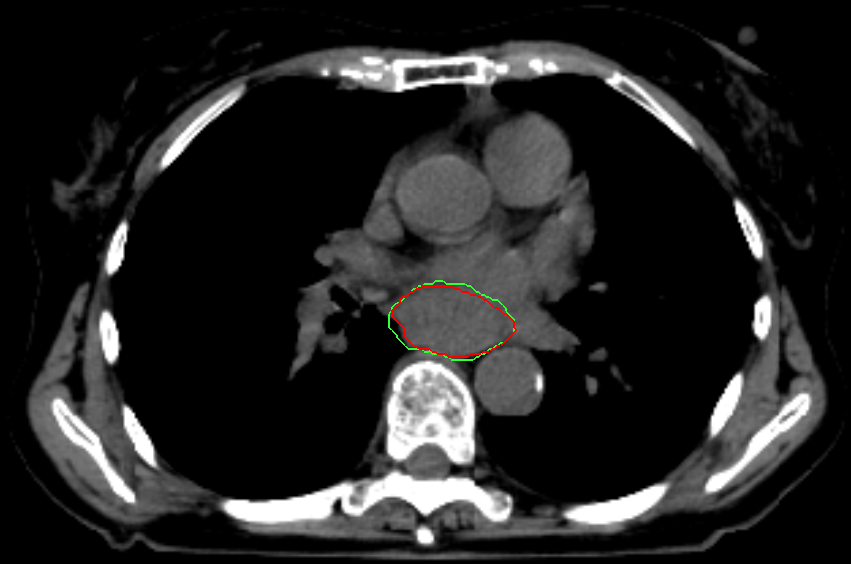}\put (-25,5) {\color{yellow}\normalsize${0.92}$}}&
{\includegraphics[width = 0.19\textwidth,trim={7cm 4.cm 5cm 4.9cm}, clip]{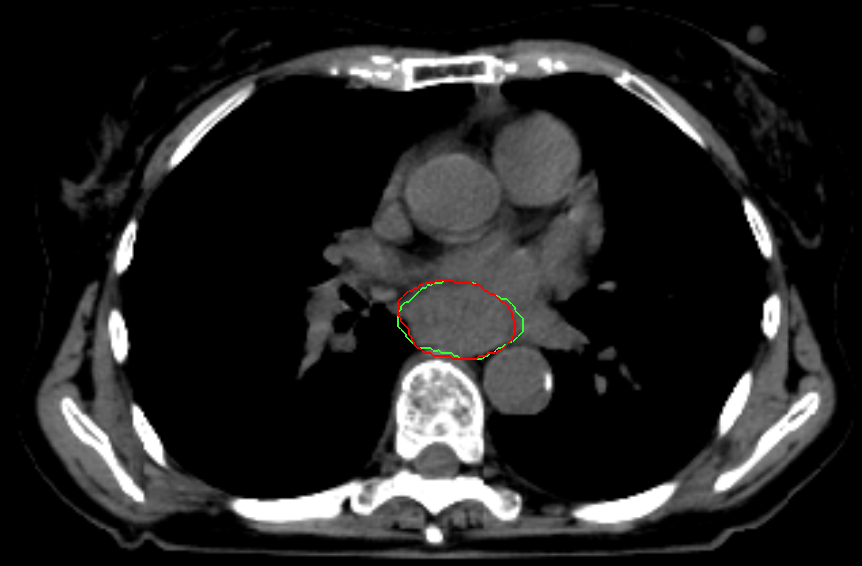}\put (-25,5) {\color{yellow}\normalsize${0.94}$}} \\
{DDAUnet-plusChA1-noChA2}  &{DDAUnet-plusChA1-noChA2}   & {DDAUnet} \\
[3pt]
\end{tabular}
\centerline{\color{gray}\rule{13cm}{0.4pt}}
\begin{tabular}{cccc}
&&&\\\includegraphics[width = 0.19\textwidth,trim={9cm 5.3cm 6cm 5.8cm}, clip]{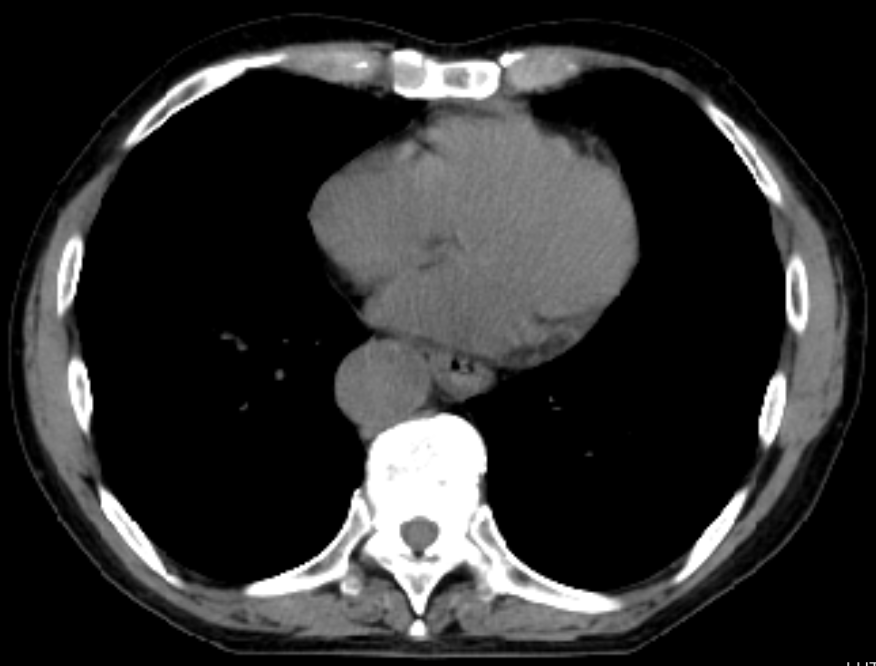}&
{\includegraphics[width = 0.19\textwidth,trim={9cm 5.2cm 6cm 5.8cm}, clip]{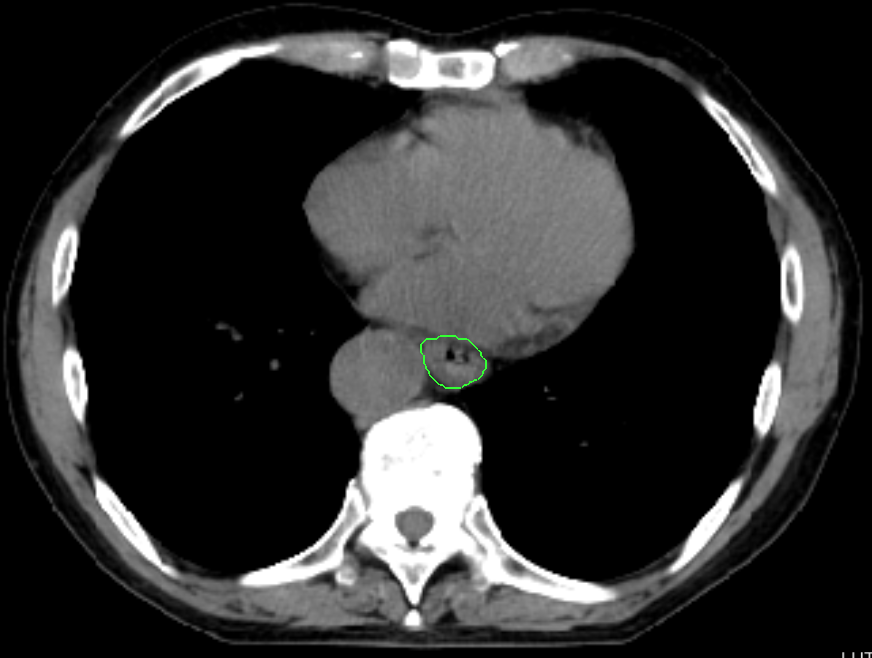}\put (-25,5) {\color{yellow}\normalsize${0.00}$}}&
{\includegraphics[width = 0.19\textwidth,trim={9cm 5.2cm 6cm 5.8cm}, clip]{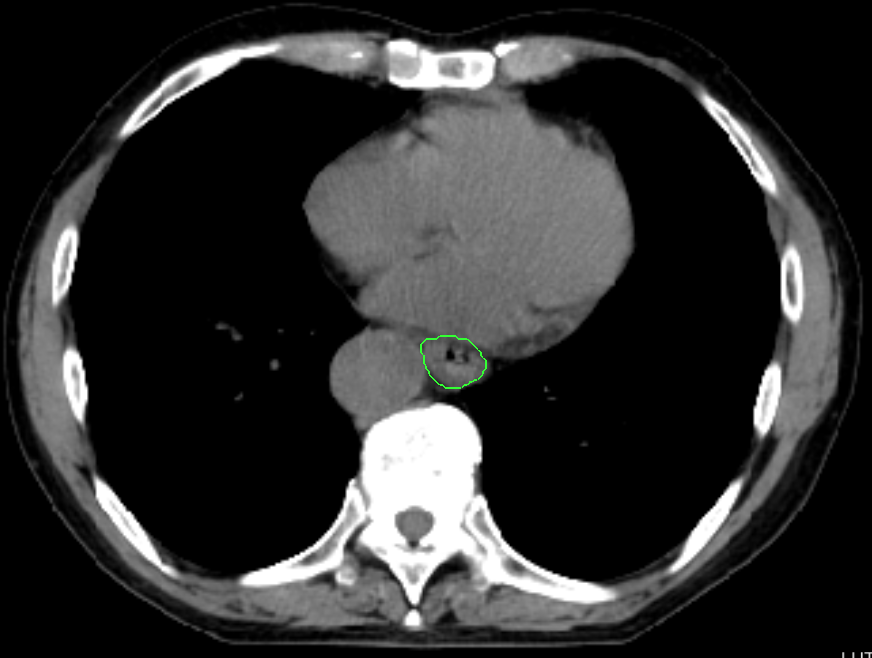}\put (-25,5) {\color{yellow}\normalsize${0.00}$}}&
{\includegraphics[width = 0.19\textwidth,trim={9cm 5.2cm 6cm 5.8cm}, clip]{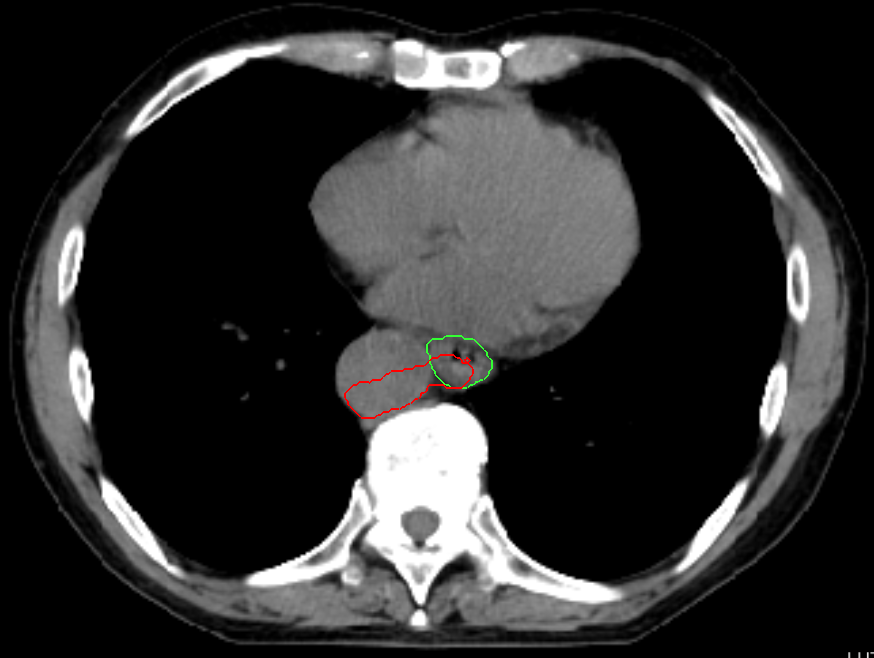}\put (-25,5) {\color{yellow}\normalsize${0.29}$}} \\
{CT scan}  & {DUnet} & {DDUnet}& {DDAUnet-noSpA-plusChA1-noChA2}  \\[0pt]
\end{tabular}
\begin{tabular}{cccc}
{\includegraphics[width = 0.19\textwidth,trim={9cm 5.2cm 6cm 6cm}, clip]{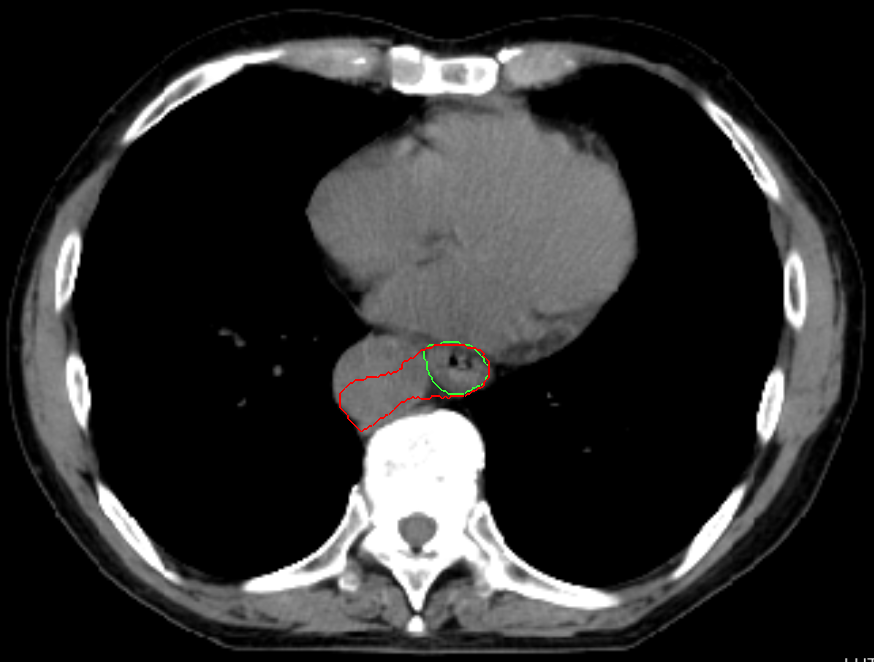}\put (-25,5) {\color{yellow}\normalsize${0.54}$}}&
{\includegraphics[width = 0.19\textwidth,trim={9cm 5.2cm 6cm 6cm}, clip]{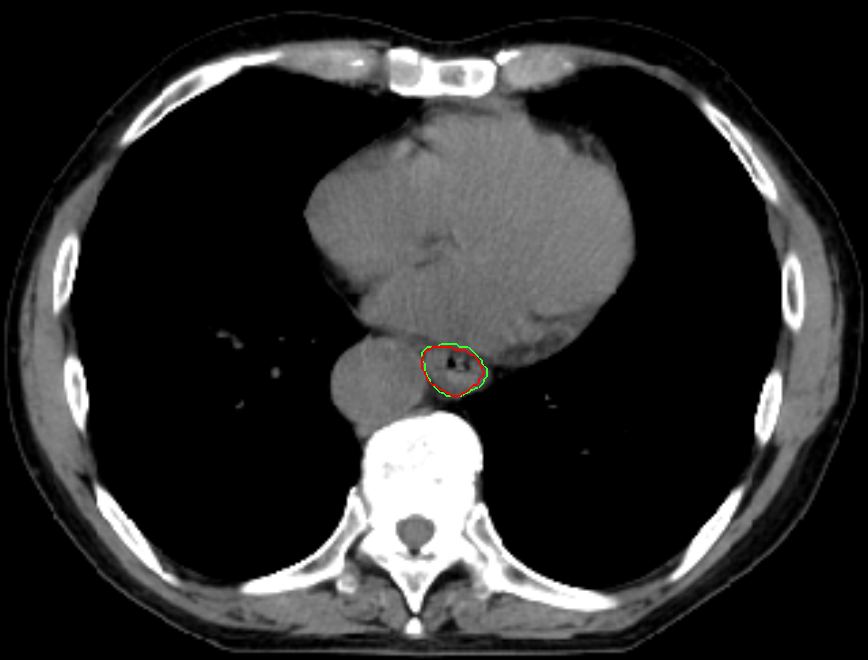}\put (-25,5) {\color{yellow}\normalsize${0.91}$}}&
{\includegraphics[width = 0.19\textwidth,trim={9cm 5.2cm 6cm 6cm}, clip]{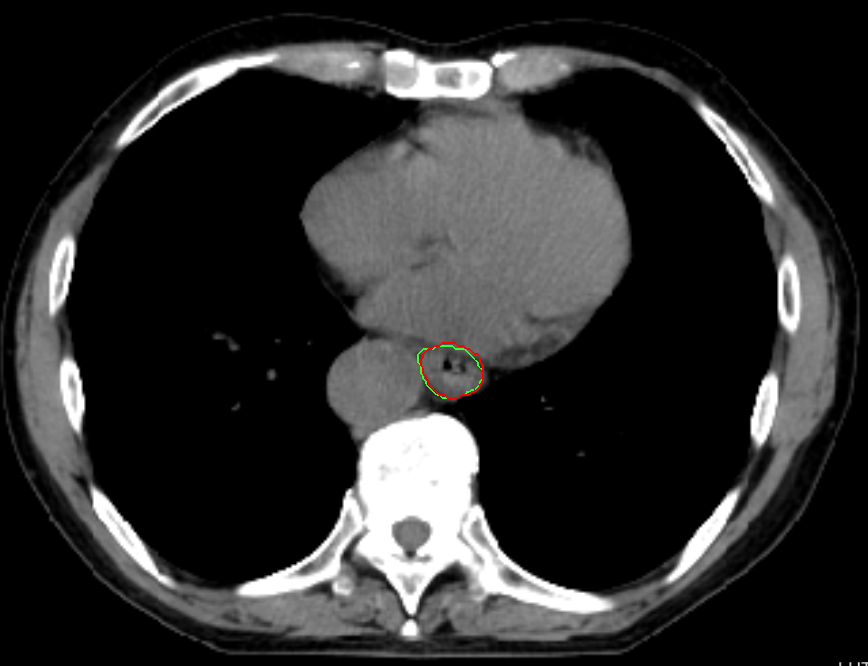}\put (-25,5) {\color{yellow}\normalsize${0.92}$}} \\
{DDAUnet-noChA2}  &{DDAUnet-plusChA1-noChA2}   & {DDAUnet} \\[3pt]
\end{tabular}
\caption{Qualitative comparison of {DDAUnet} with the other CNNs for three slices from three distinct patients. 2D DSC values are show in yellow. The manual delineation and the network results are shown by green and red contours, respectively.}
\label{fig:qualitative_comp}
\end{figure*}

As explained before, repeatability and reproducibility studies were conducted for three distinct and random splits of training and validation sets. Table \ref{table:final_results} shows the results on the independent test set after applying the largest component analysis. 
Figure \ref{fig:final_results} shows example results of the final network for some patients with different shape varieties and difficulties raised by the presence of \mbox{air pockets} or feeding tubes. The 2D $\mathrm{DSC}$ values are shown in yellow. Figure \ref{fig:qualitative_comp} shows a qualitative comparison between the different CNNs. 

\begin{figure*}[htb]
  \centering
    \includegraphics[width=19cm,clip]{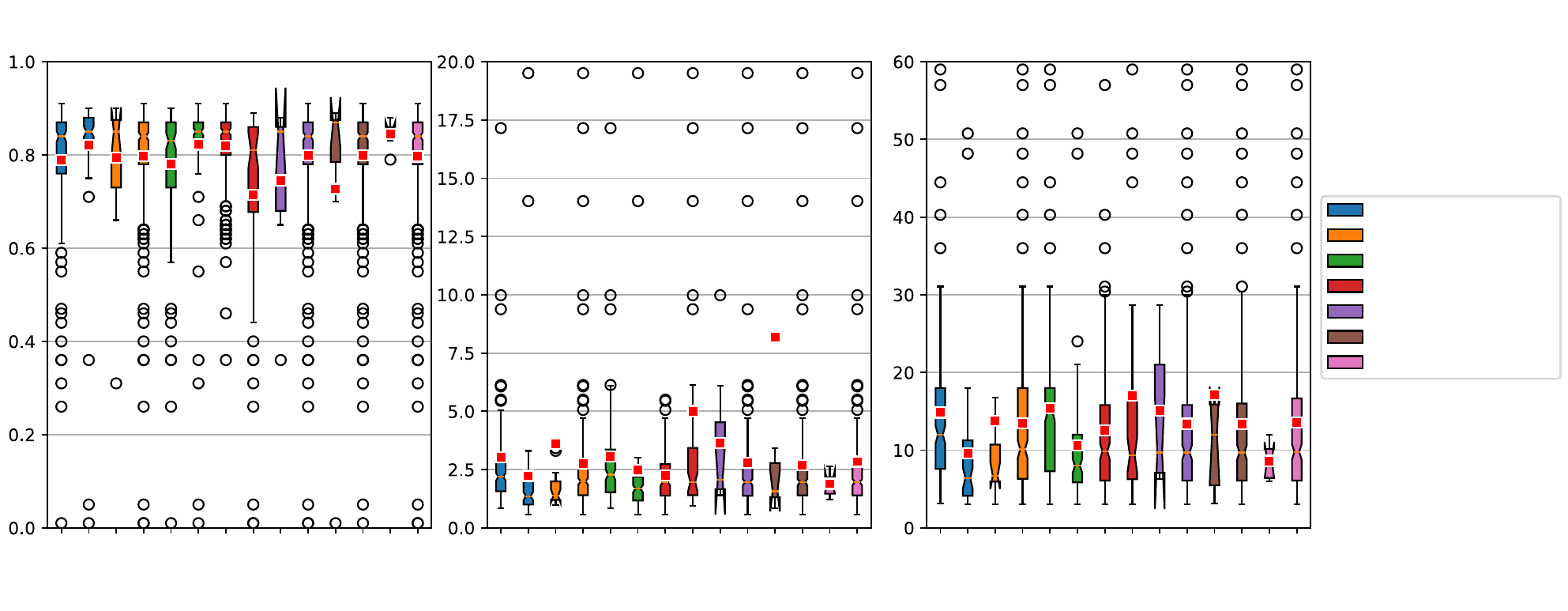}
    \put(-65,137){\footnotesize{\mbox{air pocket}}}
    \put(-65,128){\footnotesize{Feeding tube}}
    \put(-65,119){\footnotesize{Junction tumor}}
    \put(-65,110){\footnotesize{GTV>30cc}}
    \put(-65,101){\footnotesize{Hiatal hernia}}
    \put(-65,92){\footnotesize{Dislocated esoph.}}
    \put(-65,83){\footnotesize{Proximal GTV}}
    \put(-530,9){\scriptsize\rotatebox{50}{+(156)}}
    \put(-518,13){\scriptsize\rotatebox{50}{-(56)}}
    \put(-510,12){\scriptsize\rotatebox{50}{+(14)}}
    \put(-500,9){\scriptsize\rotatebox{50}{-(198)}}
    \put(-490,9){\scriptsize\rotatebox{50}{+(126)}}
    \put(-478,13){\scriptsize\rotatebox{50}{-(86)}}
    \put(-472,9){\scriptsize\rotatebox{50}{+(168)}}
    \put(-460,13){\scriptsize\rotatebox{50}{-(44)}}
    \put(-450,14){\scriptsize\rotatebox{50}{+(7)}}
    \put(-442,10){\scriptsize\rotatebox{50}{-(205)}}
    \put(-430,13){\scriptsize\rotatebox{50}{+(5)}}
    \put(-422,9){\scriptsize\rotatebox{50}{-(207)}}
    \put(-412,13){\scriptsize\rotatebox{50}{+(4)}}
    \put(-404,10){\scriptsize\rotatebox{50}{-(208)}}
    \put(-380,9){\scriptsize\rotatebox{50}{+(156)}}
    \put(-368,13){\scriptsize\rotatebox{50}{-(56)}}
    \put(-360,12){\scriptsize\rotatebox{50}{+(14)}}
    \put(-350,9){\scriptsize\rotatebox{50}{-(198)}}
    \put(-340,9){\scriptsize\rotatebox{50}{+(126)}}
    \put(-326,13){\scriptsize\rotatebox{50}{-(86)}}
    \put(-320,9){\scriptsize\rotatebox{50}{+(168)}}
    \put(-308,13){\scriptsize\rotatebox{50}{-(44)}}
    \put(-298,14){\scriptsize\rotatebox{50}{+(7)}}
    \put(-290,10){\scriptsize\rotatebox{50}{-(205)}}
    \put(-278,13){\scriptsize\rotatebox{50}{+(5)}}
    \put(-270,9){\scriptsize\rotatebox{50}{-(207)}}
    \put(-260,13){\scriptsize\rotatebox{50}{+(4)}}
    \put(-252,10){\scriptsize\rotatebox{50}{-(208)}}
    \put(-228,9){\scriptsize\rotatebox{50}{+(156)}}
    \put(-216,13){\scriptsize\rotatebox{50}{-(56)}}
    \put(-208,12){\scriptsize\rotatebox{50}{+(14)}}
    \put(-198,9){\scriptsize\rotatebox{50}{-(198)}}
    \put(-188,9){\scriptsize\rotatebox{50}{+(126)}}
    \put(-174,13){\scriptsize\rotatebox{50}{-(86)}}
    \put(-168,9){\scriptsize\rotatebox{50}{+(168)}}
    \put(-156,13){\scriptsize\rotatebox{50}{-(44)}}
    \put(-146,14){\scriptsize\rotatebox{50}{+(7)}}
    \put(-138,10){\scriptsize\rotatebox{50}{-(205)}}
    \put(-126,13){\scriptsize\rotatebox{50}{+(5)}}
    \put(-118,9){\scriptsize\rotatebox{50}{-(207)}}
    \put(-108,13){\scriptsize\rotatebox{50}{+(4)}}
    \put(-100,10){\scriptsize\rotatebox{50}{-(208)}
    \put(-380,183){\small{DSC}}
    \put(-245,183){\small{MSD (mm)}}
    \put(-97,183){\small{95\%HD (mm)}}
    }
  \caption{Results analysis for DDAUnet, $\mathrm{DSC}$, $\mathrm{MSD}$ and 95$\%\mathrm{HD}$ boxplots on the test data for different patients with or without an \mbox{air pocket}, a feeding tube in the esophagus lumen, a junction tumor, tumor volume larger than 30cc (which is defined by the median split of the GTV volumes), a hiatal hernia, a dislocated esophagus, or a proximal tumor. The outliers larger than 20 and 60 for$\mathrm{MSD}$ and 95$\%\mathrm{HD}$ have not been shown. The number of scans for each boxplot is shown in parentheses below each plot.}
  \label{fig:analytical_result} 
\end{figure*}

\begin{table*}[tb]
\centering
\caption{Results for {DDAUnet} on the independent test set, with the combined  Dice and boundary loss function.}
\begin{tabular}{c|c|c|c|c|c}
\hline
\multirow{2}{*}{Split} & DSC & CrD (mm)& CaD (mm)& MSD (mm)&95$\%\mathrm{HD}$  (mm)\\
&$\mu \pm \sigma$&$\mu \pm \sigma$&$\mu \pm \sigma$&$\mu \pm \sigma$&$\mu \pm \sigma$\\
\hline
1&0.79 $\pm$ 0.20&-6.4 $\pm$ 26.0&3.1 $\pm$ 12.8&6.2 $\pm$ 23.2&16.1 $\pm$ 28.1\\\hline
2&0.79 $\pm$ 0.19& -8.6 $\pm$ 17.7 & 4.6 $\pm$ 13.1 & 4.6 $\pm$ 16.2&14.6 $\pm$ 22.2\\\hline
3&0.78 $\pm$ 0.21&-4.6 $\pm$ 12.7&2.9 $\pm$ 10.8&5.5 $\pm$ 20.6&13.5 $\pm$ 24.3\\\hline\hline
Mean&0.79 $\pm$ 0.20&-6.5 $\pm$ 19.6&3.5 $\pm$ 12.3&5.4 $\pm$ 20.2&14.7 $\pm$ 25.0\\
\hline
\end{tabular}\label{table:final_results}
\end{table*}
In order to study the strengths and weaknesses of the final model, we manually labeled each scan with the following properties: presence of \mbox{air pockets} in the esophagus, the presence of a feeding tube in the esophageal lumen, the tumor is a junction tumor, the tumor volume is larger than 30cc (which is defined by the median split of the GTV volumes), the patient has a hiatal hernia, the tumor is in a dislocated esophagus, the tumor is located in the proximal esophagus (proximal tumor). 
Figure \ref{fig:analytical_result} shows the results of $\mathrm{DSC}$ value, $\mathrm{MSD}$ and 95$\%\mathrm{HD}$ for the mentioned tags for the final network on the test set. Results show that the network works better for patients with absence of \mbox{air pockets}, feeding tubes, or junction tumors. This may be caused by the different varieties raised by the existence of \mbox{air pockets} or foreign bodies. Also, the number of patients with a hiatal hernia, a dislocated esophagus or a proximal tumor is relatively small in the test set, not allowing to draw a conclusion.

\section{Discussion}

Esophageal GTV segmentation is not a trivial problem, due to the difficulties raised by the poor contrast with respect to its vicinity. Most research addressed only segmentation of the esophagus, while esophageal GTV segmentation has been touched in few works. Since defining a correct start and end location (slice) of the tumor in the cranial-caudal direction based on CT images alone, is not an easy task even for doctors, esophageal GTV segmentation is considered as an ill-posed problem. In this paper, for addressing the esophageal GTV segmentation we designed an efficient deep learning model. 

In terms of the training data, we collected 792 CT scans from 288 patients diagnosed with an esophageal tumor. This dataset is the largest dataset among the present works addressing esophageal tumor segmentation. Training time for a single network was in the order of 5 days. The average inference time for the final network for a cube of $255\times 255\times 255$ voxels, is $4.0 \pm 1.1$ seconds.

For tuning the proposed network, many experiments were performed in the present paper. We leveraged the DenseUnet network, already deployed in our prior work as a baseline. In order to increase the receptive field of the network, dense blocks were equipped by dilated convolutional layers, dubbed DDUnet network. Then we leveraged attention mechanisms to encourage the network to selectively filter out GTV irrelevant features. Three types of attention gates were utilized: i) a spatial attention gate in the dense blocks to filter out GTV irrelevant features in the spatial domain of each feature map, ii) a channel attention gate in the dense blocks to filter out irrelevant feature maps entirely, and iii) skip attention gates to filter out GTV irrelevant feature maps between the contracting and expanding paths of the Unet. The experiments on the validation set showed that the architecture with the spatial attention and skip attention gates, dubbed {DDAUnet}, achieved the best result. Deploying channel attention inside the dense blocks (see Figure \ref{fig:architecture}) might filter out the feature maps in early levels of the network before allowing it to extract fine features at the deeper levels. Channel attention in the skip connections filters out redundant or irrelevant feature maps during the retrieval of lost resolution. The optimized network architecture was further tuned using a large variety of loss functions, again on the validation set. Results showed that the summation of Dice and boundary loss performed best. Therefore, we introduced the {DDAUnet} with summation of Dice and boundary loss as the loss function as the final network. 

We trained the final network for three random splits of the training and validation sets. The results on the test set showed an average DSC of $0.79 \pm 0.20$, an $\mathrm{MSD}$ of $5.4 \pm 20.2 mm$, a $95\%\mathrm{HD}$ of $14.7 \pm 25.0 mm$, and cranial and caudal perpendicular distance errors of $-6.5 \pm 12.3 mm$ and $5.4 \pm 20.2 mm$, respectively. The cranial and caudal perpendicular distance errors between the ground truth and the network result show that the network overestimates at the top of the GTV by \textasciitilde 6.5 mm, and underestimates at the bottom of the GTV by \textasciitilde 3.5 mm. As slice thickness of the data was 3 mm, this translates to approximately 2 and 1 slices on average, respectively. For alleviating this issue, incorporating auxiliary information could aid the network.

Although the datasets are not comparable, in \cite{jin2019accurate} an average $\mathrm{DSC}$ score of $0.76 \pm 0.13$ was obtained on scans of 110 patients, using 5-fold cross validation. In \cite{hao2017esophagus} a $\mathrm{DSC}$ score of \mbox{0.75 $\pm$ 0.04} for four patients as the test set has been reported. In our prior work \cite{yousefi2018esophageal}, we achieved a $\mathrm{DSC}$ value of $0.73 \pm 0.20$, and a 95$\%$ mean surface distance $\mathrm{MSD}$ of $3.07 \pm 1.86$ mm for 85 CT scans from 13 distinct patients. In the work described in this paper, a higher $\mathrm{DSC}$ value was obtained.

Nowee \textit{et al.} studied the inter-observer variability in esophageal tumour delineation, and found that this variability is mainly located at the cranial and caudal border \cite{nowee2019gross}. They report a generalized conformity index for the GTV, a measure related to Dice overlap but for multiple observers, of 0.67. The human delineation variation in the cranial direction, defined as the standard deviation of the most proximal slice, was on average 9.9 mm, and 7.5 mm for the caudal direction. Although these measures are not the same as the measures reported in this paper, we cautiously conclude that the cranial and caudal error of the proposed automatic method (see Table \ref{table:final_results}) is not far from human delineation variation.

Nowee et al. also investigated the impact of incorporating FDG-PET scans in the delineation process, and concluded that although it can influence the delineated volume significantly, its impact on observer variation was limited. As a future work, we aim to study if fusion of CT with FDG-PET can aid the CNNs to improve the extracted features and subsequently the segmentation results. 

For a close inspection, we investigated the results on the independent test set for the final network. We labelled the patients in the test set with different tags, including the presence of \mbox{air pockets}, feeding tube, junction tumor, tumor volume $>30cc$, hiatal hernia, dislocated esophagus, proximal tumor. Inspection of the final results (see Figure \ref{fig:analytical_result}) showed that the network performed better for patients with an absence of \mbox{air pockets}, feeding tubes in the esophagus lumen, or junction tumors. A lower performance was obtained for smaller tumors ($<30cc$), while the strength of the network for patients with a dislocated esophagus, a proximal tumor or a hiatal hernia was not judge-able. Therefore, enriching the dataset with more patients with the mentioned properties would potentially improve the performance of the model. Also, incorporating endoscopic findings in the process of segmentation can be considered as a future work to investigate if that can aid CNNs to reduce errors specially at the start and end of the GTV.

\section{Conclusion}

In this study, we collected a large set of CT scans from 288 distinct patients with esophageal cancer. To the best of our knowledge this is the largest dataset in esophageal tumor segmentation literature to date. We showed that despite the difficulties raised by poor contrast of esophageal tumors with respect to their neighbouring tissues, varieties in shape and location of tumor, presence of \mbox{air pockets} and foreign bodies, the proposed method, dubbed dilated dense attention Unet (DDAUnet), could segment the gross tumor volume with a mean surface distance of $5.4 \pm 20.2mm$. 

\section{Acknowledgments}
Femke~P.~Peters is acknowledged for delineation of the data. 
\bibliographystyle{unsrt}
\bibliography{references}

\begin{thebibliography}{10}

\bibitem{enzinger2003esophageal}
Peter~C Enzinger and Robert~J Mayer.
\newblock Esophageal cancer.
\newblock {\em New England Journal of Medicine}, 349(23):2241--2252, 2003.

\bibitem{ferlay2015cancer}
Jacques Ferlay, Isabelle Soerjomataram, Rajesh Dikshit, Sultan Eser, Colin
  Mathers, Marise Rebelo, Donald Maxwell~Parkin, David Forman, and Freddie
  Bray.
\newblock Cancer incidence and mortality worldwide: sources, methods and major
  patterns in globocan 2012.
\newblock {\em International Journal of Cancer}, 136(5):E359--E386, 2015.

\bibitem{mamede2007fdg}
Marcelo Mamede, Paula Abreu-e Lima, Maria~Raquel Oliva, V{\^a}nia Nos{\'e},
  Harvey Mamon, and Victor~H Gerbaudo.
\newblock {FDG-PET/CT} tumor segmentation-derived indices of metabolic activity
  to assess response to neoadjuvant therapy and progression-free survival in
  esophageal cancer: correlation with histopathology results.
\newblock {\em American Journal of Clinical Oncology}, 30(4):377--388, 2007.

\bibitem{nowee2019gross}
Marlies~E Nowee, Francine~EM Voncken, Alexis~NTJ Kotte, Lucas Goense, Peter~SN
  van Rossum, Astrid~LHMW van Lier, Stijn~W Heijmink, Berthe~MP Aleman, et~al.
\newblock Gross tumour delineation on computed tomography and positron emission
  tomography-computed tomography in oesophageal cancer: A nationwide study.
\newblock {\em Clinical and Translational Radiation Oncology}, 14:33--39, 2019.

\bibitem{charles2009esophageal}
Thomas~R Charles, John~G Hunter, and Blair~AA Jobe.
\newblock {\em Esophageal cancer: principles and practice}.
\newblock Demos Medical Publishing, 2009.

\bibitem{rousson2006probabilistic}
Mikael Rousson, Ying Bai, Chenyang Xu, and Frank Sauer.
\newblock Probabilistic minimal path for automated esophagus segmentation.
\newblock In {\em Medical Imaging 2006: Image Processing}, volume 6144, page
  614449. International Society for Optics and Photonics, 2006.

\bibitem{lever2014quantification}
Frederiek~M Lever, Irene~M Lips, Sjoerd~PM Crijns, Onne Reerink, Astrid~LHMW
  van Lier, Marinus~A Moerland, Marco van Vulpen, and Gert~J Meijer.
\newblock Quantification of esophageal tumor motion on cine-magnetic resonance
  imaging.
\newblock {\em International Journal of Radiation Oncology* Biology* Physics},
  88(2):419--424, 2014.

\bibitem{jin2019deep}
Dakai Jin, Dazhou Guo, Tsung-Ying Ho, Adam~P Harrison, Jing Xiao, Chen-kan
  Tseng, and Le~Lu.
\newblock Deep esophageal clinical target volume delineation using encoded {3D}
  spatial context of tumors, lymph nodes, and organs at risk.
\newblock In {\em MICCAI}, pages 603--612. Springer, 2019.

\bibitem{xue2017fully}
Di-Xiu Xue, Rong Zhang, Yuan-Yuan Zhao, Jian-Ming Xu, and Ya-Lei Wang.
\newblock Fully convolutional networks with double-label for esophageal cancer
  image segmentation by self-transfer learning.
\newblock In {\em Ninth International Conference on Digital Image Processing
  (ICDIP 2017)}, volume 10420, page 104202D. International Society for Optics
  and Photonics, 2017.

\bibitem{liang2020auto}
Ying Liang, Diane Schott, Ying Zhang, Zhiwu Wang, Haidy Nasief, Eric Paulson,
  William Hall, Paul Knechtges, Beth Erickson, and X~Allen Li.
\newblock Auto-segmentation of pancreatic tumor in multi-parametric {MRI} using
  deep convolutional neural networks.
\newblock {\em Radiotherapy and Oncology}, 145:193--200, 2020.

\bibitem{yousefi2018esophageal}
Sahar Yousefi, Hessam Sokooti, Mohamed~S Elmahdy, Femke~P Peters, Mohammad
  T~Manzuri Shalmani, Roel~T Zinkstok, and Marius Staring.
\newblock Esophageal gross tumor volume segmentation using a {3D} convolutional
  neural network.
\newblock In {\em MICCAI}, pages 343--351. Springer, 2018.

\bibitem{huang2017densely}
Gao Huang, Zhuang Liu, Kilian~Q Weinberger, and Laurens van~der Maaten.
\newblock Densely connected convolutional networks.
\newblock In {\em CVPR}, pages 4700--4708, 2017.

\bibitem{fieselmann2008automatic}
Andreas Fieselmann, Stefan Lautenschl{\"a}ger, Frank Deinzer, and Bj{\"o}rn
  Poppe.
\newblock Automatic detection of air holes inside the esophagus in {CT} images.
\newblock In {\em Bildverarbeitung f{\"u}r die Medizin 2008}, pages 397--401.
  Springer, 2008.

\bibitem{fieselmann2008esophagus}
Andreas Fieselmann, Stefan Lautenschl{\"a}ger, Frank Deinzer, Matthias John,
  and Bj{\"o}rn Poppe.
\newblock Esophagus segmentation by spatially-constrained shape interpolation.
\newblock In {\em Bildverarbeitung f{\"u}r die Medizin 2008}, pages 247--251.
  Springer, 2008.

\bibitem{feulner2011probabilistic}
Johannes Feulner, S~Kevin Zhou, Matthias Hammon, Sascha Seifert, Martin Huber,
  Dorin Comaniciu, Joachim Hornegger, and Alexander Cavallaro.
\newblock A probabilistic model for automatic segmentation of the esophagus in
  {3-D} {CT} scans.
\newblock {\em IEEE Transactions on Medical Imaging}, 30(6):1252--1264, 2011.

\bibitem{huang2006semi}
Tzung-Chi Huang, Geoffrey Zhang, Thomas Guerrero, George Starkschall, Kan-Ping
  Lin, and Ken Forster.
\newblock Semi-automated {CT} segmentation using optic flow and fourier
  interpolation techniques.
\newblock {\em Computer Methods and Programs in Biomedicine}, 84(2-3):124--134,
  2006.

\bibitem{feulner2009fast}
Johannes Feulner, S~Kevin Zhou, Alexander Cavallaro, Sascha Seifert, Joachim
  Hornegger, and Dorin Comaniciu.
\newblock Fast automatic segmentation of the esophagus from {3D} {CT} data
  using a probabilistic model.
\newblock In {\em MICCAI}, pages 255--262. Springer, 2009.

\bibitem{feulner2010model}
Johannes Feulner, S~Kevin Zhou, Martin Huber, Alexander Cavallaro, Joachim
  Hornegger, and Dorin Comaniciu.
\newblock Model-based esophagus segmentation from {CT} scans using a spatial
  probability map.
\newblock In {\em MICCAI}, pages 95--102. Springer, 2010.

\bibitem{kurugol2010locally}
Sila Kurugol, Necmiye Ozay, Jennifer~G Dy, Gregory~C Sharp, and Dana~H Brooks.
\newblock Locally deformable shape model to improve {3D} level set based
  esophagus segmentation.
\newblock In {\em ICPR 2010}, pages 3955--3958. IEEE, 2010.

\bibitem{kurugol2011centerline}
Sila Kurugol, Erhan Bas, Deniz Erdogmus, Jennifer~G Dy, Gregory~C Sharp, and
  Dana~H Brooks.
\newblock Centerline extraction with principal curve tracing to improve {3D}
  level set esophagus segmentation in {CT} images.
\newblock In {\em EMBC 2011}, pages 3403--3406. IEEE, 2011.

\bibitem{yang2017atlas}
Jinzhong Yang, Benjamin Haas, Raymond Fang, Beth~M Beadle, Adam~S Garden,
  Zhongxing Liao, Lifei Zhang, Peter Balter, et~al.
\newblock Atlas ranking and selection for automatic segmentation of the
  esophagus from {CT} scans.
\newblock {\em Physics in Medicine \& Biology}, 62(23):9140, 2017.

\bibitem{yousefi2021asl}
Sahar Yousefi, Hessam Sokooti, Wouter~M Teeuwisse, Dennis~FR Heijtel, Aart~J
  Nederveen, Marius Staring, and Matthias~JP van Osch.
\newblock {ASL} to {PET} translation by a semi-supervised residual-based
  attention-guided convolutional neural network.
\newblock {\em arXiv preprint arXiv:2103.05116}, 2021.

\bibitem{pezzotti2020adaptive}
Nicola Pezzotti, Sahar Yousefi, Mohamed~S Elmahdy, Jeroen van Gemert,
  Christophe Sch{\"u}lke, Mariya Doneva, Tim Nielsen, Sergey Kastryulin,
  Boudewijn~PF Lelieveldt, Matthias~JP van Osch, Elwin de~Weerdt, and Marius
  Staring.
\newblock An adaptive intelligence algorithm for undersampled knee {MRI}
  reconstruction.
\newblock {\em IEEE Access}, pages 204825--204838, 2020.

\bibitem{yousefi2019fast}
Sahar Yousefi, Lydiane Hirschler, Merlijn van~der Plas, Mohamed~S Elmahdy,
  Hessam Sokooti, Matthias Van~Osch, and Marius Staring.
\newblock Fast dynamic perfusion and angiography reconstruction using an
  end-to-end {3D} convolutional neural network.
\newblock In {\em MLMRI}, pages 25--35. Springer, 2019.

\bibitem{elmahdy2018evaluation}
Mohamed~S Elmahdy, Thyrza Jagt, Sahar Yousefi, Hessam Sokooti, Roel Zinkstok,
  Mischa Hoogeman, and Marius Staring.
\newblock Evaluation of multi-metric registration for online adaptive proton
  therapy of prostate cancer.
\newblock In {\em International Workshop on Biomedical Image Registration},
  pages 94--104. Springer, 2018.

\bibitem{elmahdy2019robust}
Mohamed~S Elmahdy, Thyrza Jagt, Roel~Th Zinkstok, Yuchuan Qiao, Rahil Shahzad,
  Hessam Sokooti, Sahar Yousefi, Luca Incrocci, CAM Marijnen, Mischa Hoogeman,
  et~al.
\newblock Robust contour propagation using deep learning and image registration
  for online adaptive proton therapy of prostate cancer.
\newblock {\em Medical physics}, 46(8):3329--3343, 2019.

\bibitem{fechter20173d}
Tobias Fechter, Sonja Adebahr, Dimos Baltas, Ismail~Ben Ayed, Christian
  Desrosiers, and Jose Dolz.
\newblock A {3D} fully convolutional neural network and a random walker to
  segment the esophagus in {CT}.
\newblock {\em Journal of Medical Physics}, 2017.

\bibitem{trullo2017fully}
Roger Trullo, Caroline Petitjean, Dong Nie, Dinggang Shen, and Su~Ruan.
\newblock Fully automated esophagus segmentation with a hierarchical deep
  learning approach.
\newblock In {\em ICSIPA 2017}, pages 503--506. IEEE, 2017.

\bibitem{hao2017esophagus}
Zhaojun Hao, Jiwei Liu, and Jianfei Liu.
\newblock Esophagus tumor segmentation using fully convolutional neural network
  and graph cut.
\newblock In {\em Chinese Intelligent Systems Conference}, pages 413--420.
  Springer, 2017.

\bibitem{jin2019accurate}
Dakai Jin, Dazhou Guo, Tsung-Ying Ho, Adam~P Harrison, Jing Xiao, Chen-kan
  Tseng, and Le~Lu.
\newblock Accurate esophageal gross tumor volume segmentation in {PET/CT} using
  two-stream chained {3D} deep network fusion.
\newblock In {\em MICCAI}, pages 182--191. Springer, 2019.

\bibitem{cordts2016cityscapes}
Marius Cordts, Mohamed Omran, Sebastian Ramos, Timo Rehfeld, Markus Enzweiler,
  Rodrigo Benenson, Uwe Franke, Stefan Roth, and Bernt Schiele.
\newblock The cityscapes dataset for semantic urban scene understanding.
\newblock In {\em CVPR}, pages 3213--3223, 2016.

\bibitem{simonyan2014very}
Karen Simonyan and Andrew Zisserman.
\newblock Very deep convolutional networks for large-scale image recognition.
\newblock {\em ICLR}, 2014.

\bibitem{milletari2016v}
Fausto Milletari, Nassir Navab, and Seyed-Ahmad Ahmadi.
\newblock V-net: Fully convolutional neural networks for volumetric medical
  image segmentation.
\newblock In {\em 3DV 2016}, pages 565--571. IEEE, 2016.

\bibitem{kervadec2019boundary}
Hoel Kervadec, Jihene Bouchtiba, Christian Desrosiers, Eric Granger, Jose Dolz,
  and Ismail~Ben Ayed.
\newblock Boundary loss for highly unbalanced segmentation.
\newblock In {\em International Conference on Medical imaging with deep
  learning}, pages 285--296, 2019.

\bibitem{caliva2019distance}
Francesco Caliva, Claudia Iriondo, Alejandro~Morales Martinez, Sharmila
  Majumdar, and Valentina Pedoia.
\newblock Distance map loss penalty term for semantic segmentation.
\newblock {\em MIDL}, 2019.

\bibitem{wang2018focal}
Pei Wang and Albert~CS Chung.
\newblock Focal {Dice} loss and image dilation for brain tumor segmentation.
\newblock In {\em Deep Learning in Medical Image Analysis and Multimodal
  Learning for Clinical Decision Support}, pages 119--127. Springer, 2018.

\end{thebibliography}

\ifCLASSOPTIONcaptionsoff
  \newpage
\fi

\end{document}